\documentclass[twocolumn,english,superscriptaddress]{revtex4}
\usepackage[T1]{fontenc}
\usepackage[latin9]{inputenc}
\usepackage{color}
\usepackage{amsmath}
\usepackage{amssymb}
\usepackage{stmaryrd}
\usepackage{graphicx}
\usepackage{esint}

\makeatletter

\providecommand{\tabularnewline}{\\}

\@ifundefined{textcolor}{}
{%
 \definecolor{BLACK}{gray}{0}
 \definecolor{WHITE}{gray}{1}
 \definecolor{RED}{rgb}{1,0,0}
 \definecolor{GREEN}{rgb}{0,1,0}
 \definecolor{BLUE}{rgb}{0,0,1}
 \definecolor{CYAN}{cmyk}{1,0,0,0}
 \definecolor{MAGENTA}{cmyk}{0,1,0,0}
 \definecolor{YELLOW}{cmyk}{0,0,1,0}
}

\usepackage{babel}

\usepackage{babel}

\usepackage{babel}

\usepackage{babel}

\usepackage{babel}

\usepackage{babel}

\usepackage{babel}

\usepackage{babel}

\usepackage{babel}

\usepackage{babel}

\makeatother

\usepackage{babel}
\begin{document}

\title{Vestigial chiral and charge orders from bidirectional spin-density waves: Application to the iron-based superconductors}

\author{R. M. Fernandes}

\affiliation{School of Physics and Astronomy, University of Minnesota, Minneapolis
55455, USA}

\author{S. A. Kivelson}

\affiliation{Department of Physics, Stanford University, Stanford, California
94305, USA}

\author{E. Berg}

\affiliation{Department of Condensed Matter Physics, Weizmann Institute of Science,
Rehovot, Israel 76100}
\begin{abstract}
Recent experiments in optimally hole-doped iron arsenides have revealed
a novel magnetically ordered ground state that preserves tetragonal
symmetry, consistent with either a charge-spin density wave (CSDW),
which displays a non-uniform magnetization, or a spin-vortex crystal
(SVC), which displays a non-collinear magnetization. Here we show
that, similarly to the partial melting of the usual stripe antiferromagnet
into a nematic phase, either of these phases can also melt in two
stages. As a result, intermediate paramagnetic phases with vestigial
order appears: a checkerboard charge density-wave for the CSDW ground
state, characterized by an Ising-like order parameter, and a remarkable
spin-vorticity density-wave for the SVC ground state -- a triplet
d-density wave characterized by a vector chiral order parameter. We
propose experimentally detectable signatures of these phases, show
that their fluctuations can enhance the superconducting transition
temperature, and discuss their relevance to other correlated materials. 
\end{abstract}
\maketitle

\section{Introduction}

\begin{figure}
\begin{centering}
\includegraphics[width=1\columnwidth]{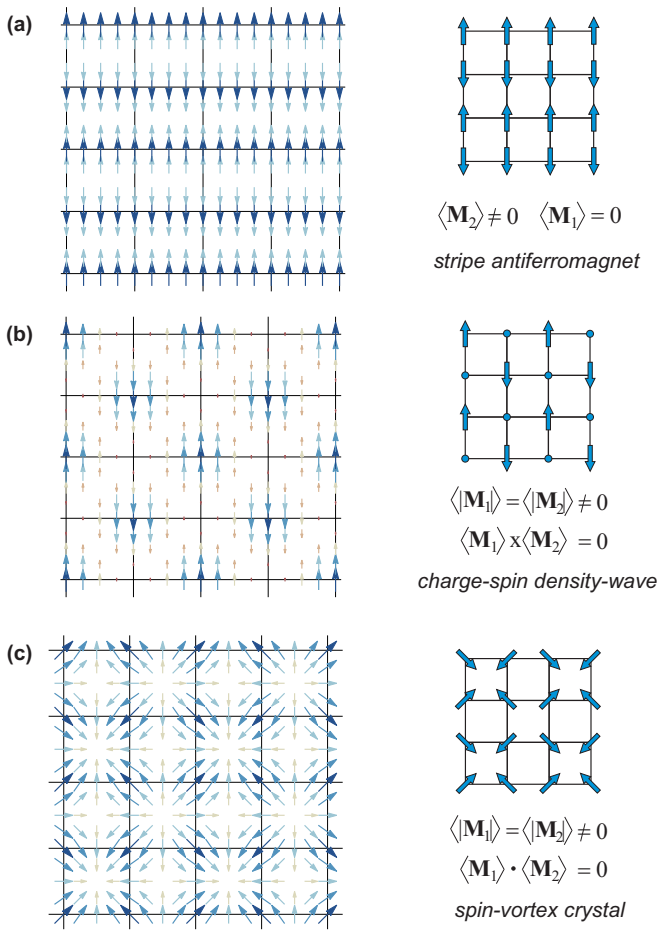} 
\par\end{centering}

\protect\protect\protect\protect\protect\caption{Magnetic ground states of the iron pnictides: (a) stripe antiferromagnet,
(b) charge-spin density-wave (CSDW), (c) spin-vortex crystal (SVC).
The first is orthorhombic with a doubled unit cell; the latter two
remain tetragonal but with a quadrupled unit cell. $\mathbf{M}_{1}$
and $\mathbf{M}_{2}$ are the magnetic order parameters corresponding
to the ordering vectors $\mathbf{Q}_{1}=\left(\pi,0\right)$ and $\mathbf{Q}_{2}=\left(0,\pi\right)$.
The left panels are the actual spin density-wave patterns in real
space, whereas the right panels are schematic representations focusing
on the magnetization at the lattice sites.}

\label{fig_magnetic_gs} 
\end{figure}

One of the hallmarks of the superconducting state of the iron-based
materials \cite{reviews} is its typical proximity to a stripe magnetically
ordered state, with spins aligned parallel to each other along one
in-plane direction and anti-parallel along the other (see Fig. \ref{fig_magnetic_gs}a)
\cite{Dagotto12}. As a result, this stripe state breaks two distinct
symmetries of the high-temperature paramagnetic-tetragonal state:
a continuous spin-rotational $O(3)$ symmetry and an Ising-like $Z_{2}$
symmetry related to the equivalence of the $x$ and $y$ directions
\cite{Fang08,Xu08,Johannes09,Eremin10,Abrahams11,Fernandes12,Dagotto14}.
Magnetic fluctuations present in the paramagnetic state can cause
these two symmetries to be broken at different temperatures, giving
rise to an intermediate nematic phase that preserves the spin-rotational
$O(3)$ symmetry but, as a ``vestige'' of the stripe order \cite{Kivelson14},
breaks the tetragonal $Z_{2}$ symmetry \cite{Fernandes14}. Indeed,
in the phase diagrams of most iron-based superconductors, the magnetic
transition line is closely followed by the structural/nematic one
at slightly higher temperatures. The corresponding nematic degrees
of freedom impact not only the normal state electronic properties
\cite{Fisher10,Davis10,ZXshen11,Fisher12,Matsuda12,Gallais13,Dai14,Rosenthal14,w_ku10,Devereaux12}
but also the onset and gap structure of the superconducting state
\cite{Fernandes_Millis,Lederer14,Kang14}.

Recently, experiments in the hole-doped pnictides $\text{Ba}(\text{Fe}_{1-x}\text{Mn}_{x})_{2}\text{As}_{2}$
\cite{Kim10}, $(\text{Ba}_{1-x}\text{Na}_{x})\text{Fe}_{2}\text{As}_{2}$
\cite{Avci14}, and $(\text{Ba}_{1-x}\text{K}_{x})\text{Fe}_{2}\text{As}_{2}$
\cite{Bohmer14} have revealed another type of magnetically ordered
state that does not break the tetragonal $Z_{2}$ symmetry of the
lattice. Neutron scattering experiments \cite{Kim10,Avci14} showed
that its magnetic Bragg peaks are at the same momenta as in the stripe
magnetic phase -- namely, $\mathbf{Q}_{1}=\left(\pi,0\right)$ and
$\mathbf{Q}_{2}=\left(0,\pi\right)$ in the Fe-only Brillouin zone.
Consequently, it has been proposed \cite{Avci14,xiaoyu14,Wang_arxiv_14,Kang15,Gastiasoro15}
that the tetragonal magnetic state is the realization of one of two
possible biaxial (i.e. double-$\mathbf{Q}$) magnetic orders \cite{Eremin10,Lorenzana08,giovannetti,Brydon11}.
One possibility is a ``charge-spin density wave'' (CSDW), displaying
a non-uniform magnetization which vanishes at the even lattice sites
and is staggered along the odd lattice sites (Fig. \ref{fig_magnetic_gs}b).
The other option is a ``spin-vortex crystal'' (SVC), in which the
magnetization is non-collinear (but coplanar) and forms spin vortices
staggered across the plaquettes (Fig. \ref{fig_magnetic_gs}c). Both
CSDW and SVC phases are tetragonal, but have a unit cell four times
larger than the paramagnetic phase. Interestingly, in $(\text{Ba}_{1-x}\text{Na}_{x})\text{Fe}_{2}\text{As}_{2}$
and $(\text{Ba}_{1-x}\text{K}_{x})\text{Fe}_{2}\text{As}_{2}$, the
tetragonal magnetic state is observed very close to optimal doping
\cite{Avci14,Bohmer14}, where superconductivity displays its highest
transition temperature. Therefore, understanding the properties of
these biaxial tetragonal magnetic phases is important to assess their
relevance for the superconductivity.

In this paper, we show that both the CSDW and the SVC magnetic phases
support composite order parameters that can condense at temperatures
above the onset of magnetic order, and whose fluctuations can help
enhancing $T_{c}$. As with the nematic phase, these partially ordered
phases are paramagnetic, i.e. fluctuations restore the time-reversal
symmetry that is broken in the ground state. In contrast to the nematic
phase, however, they preserve the point group symmetry of the lattice,
but break other symmetries, including translational symmetry \cite{Chern12}.
In particular, upon melting the CSDW phase, we find a vestigial \emph{Ising-like}
charge-density wave (CDW) phase with ordering vector $\mathbf{Q}_{1}+\mathbf{Q}_{2}=\left(\pi,\pi\right)$,
in which the previously magnetized sites acquire a different charge
than the previously non-magnetized sites. On the other hand, upon
melting the SVC ground state, we find a vestigial phase that retains
memory of the preferred plane of magnetization (in spin space), and
of the staggering of the spin vortices across the plaquettes. This
spin-vorticity density-wave (SVDW) is a triplet d-density wave characterized
by a \emph{vector chiral} order parameter, which is manifested as
a spin-current density-wave with modulation $\mathbf{Q}_{1}+\mathbf{Q}_{2}=\left(\pi,\pi\right)$.
Besides shedding light on the magnetism of hole-doped iron pnictides,
our results provide a novel microscopic mechanism for the formation
of \emph{d}-density waves, which have also been proposed in cuprates
\cite{Chakravarty01} and heavy fermions \cite{Chakravarty14}.

The paper is organized as follows: in Section \ref{sec_Theoretical_model}
we present the theoretical model that gives rise to the vestigial
CDW and SVDW orders. Section \ref{sec_Microscopics} discusses the
implications of these vestigial orders for both the normal state and
superconducting state properties. Concluding remarks are presented
in Section \ref{sec_Concluding_remarks}. To make the paper transparent
and accessible, all formal details are presented in appendices. Appendix
\ref{Appendix_saddle_point} contains the derivation of the saddle-point
equations that give the phase diagram of the SVDW phase discussed
in Section \ref{sec_Theoretical_model}. In Appendix \ref{Appendix_free_energy}
we derive microscopically the free energy discussed in Section \ref{sec_Microscopics}.
Finally, Appendix \ref{Appendix_pairing} presents the derivation
of the effective pairing interactions promoted by CDW and SVDW fluctuations
discussed in Section \ref{sec_Microscopics}.

\section{Theoretical model for the vestigial phases \label{sec_Theoretical_model}}

\subsection{Effective action}

We define two magnetic order parameters, $\mathbf{M}_{1}$ and $\mathbf{M}_{2}$,
associated with the two ordering vectors $\mathbf{Q}_{1}=\left(\pi,0\right)$
and $\mathbf{Q}_{2}=\left(0,\pi\right)$, respectively. Thus, the
local spin is given by $\mathbf{S}\left(\mathbf{r}\right)=\sum_{i}\mathbf{M}_{i}\mathrm{e}^{i\mathbf{Q}_{i}\cdot\mathbf{r}}$.
As discussed in Refs. \cite{Eremin10,Fernandes12,xiaoyu14,Wang_arxiv_14,Kang15,Lorenzana08,giovannetti,Brydon11},
the most general lowest order action that respects the tetragonal
and spin-rotational symmetries is given by: 
\begin{align}
\mathcal{S}\left[\mathbf{M}_{i}\right] & =\int_{q}\chi_{q}^{-1}\left(\mathbf{M}_{1}^{2}+\mathbf{M}_{2}^{2}\right)+\frac{u}{2}\int_{x}\left(\mathbf{M}_{1}^{2}+\mathbf{M}_{2}^{2}\right)^{2}\nonumber \\
 & -\frac{g}{2}\int_{x}\left(\mathbf{M}_{1}^{2}-\mathbf{M}_{2}^{2}\right)^{2}+2w\int_{x}\left(\mathbf{M}_{1}\cdot\mathbf{M}_{2}\right)^{2}\ .\label{action}
\end{align}

For simplicity, we will consider the finite temperature problem, but
the same conclusions can be extended to the quantum case. Here, $\int_{q}\equiv\int\frac{d^{d}q}{\left(2\pi\right)^{d}}$
and $\int_{x}\equiv\int d^{d}x$ where $\mathbf{q}$ is the momentum
and $\mathbf{x}$ is the position. In the neighborhood of a finite
$T$ magnetic transition, and for a quasi-2D system, we can use the
small $q$ expansion $\chi_{q}^{-1}\approx r_{0}+q_{\parallel}^{2}+J_{z}\sin^{2}\frac{q_{z}}{2}$,
where $r_{0}$ is the distance to the mean-field magnetic critical
point.

\begin{figure}
\begin{centering}
\includegraphics[width=0.99\columnwidth]{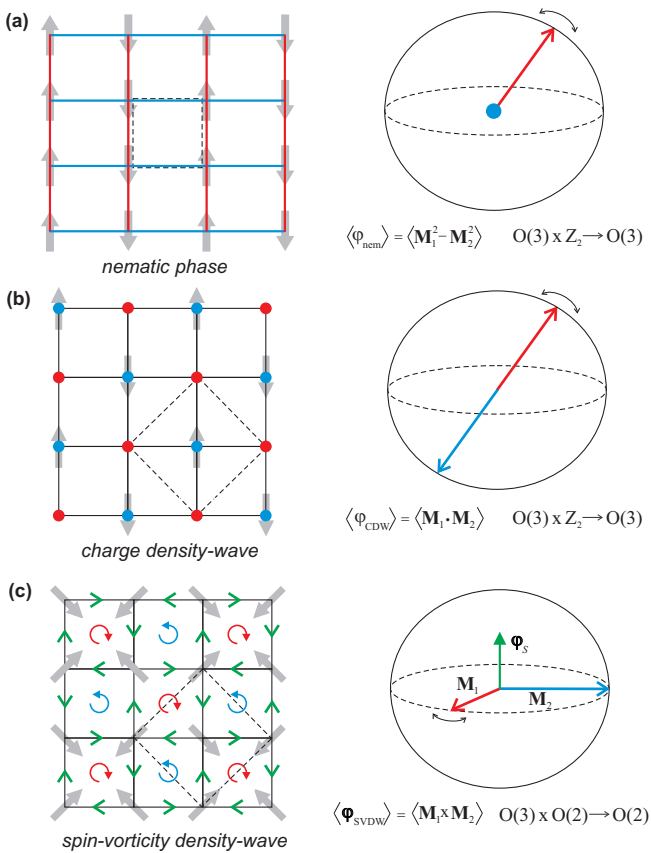} 
\par\end{centering}

\protect\protect\protect\protect\protect\caption{The vestigial composite states associated with (a) the stripe antiferromaget,
(b) the CSDW state, and (c) the SVC state. The real-space spins (in
gray) and the magnetic order parameters in spin space (red and blue
arrows, representing $\mathbf{M}_{1}$ and $\mathbf{M}_{2}$, respectively)
should be understood as fluctuating, i.e. $\left\langle \mathbf{M}_{i}\right\rangle =0$
in all cases. In (a), the vestigial state is nematic (unequal blue
and red bonds), associated with selecting between $\mathbf{M}_{1}$
and $\mathbf{M}_{2}$ fluctuations in spin space. The original unit
cell is shown as a dashed square. In (b), the vestigial state breaks
translational symmetry via a checkerboard charge density-wave (unequal
blue and red sites). $\mathbf{M}_{1}$ and $\mathbf{M}_{2}$ are locked
to be collinear in spin space. In (c), the vestigial spin-vorticity
density-wave state breaks inversion and translational symmetries via
a staggered pattern of spin vortices in the center of the plaquettes
(unequal blue and red plaquettes). The corresponding spin-current
pattern is shown by the green arrows. Because $\mathbf{M}_{1}$ and
$\mathbf{M}_{2}$ are locked to be orthogonal in spin space, the residual
spin-rotational symmetry is $O(2)$ instead of $O(3)$. Both (b) and
(c) preserve tetragonal symmetry, as shown by the dashed-line unit
cell. \label{fig_emergent_states}}
\end{figure}

The quartic coefficients $u$, $g$, $w$ determine the nature of
the magnetic ground state. These are, in turn, sensitive to microscopic
considerations. The localized $J_{1}$-$J_{2}$ model favors positive
$g$ and $w$~\cite{chandra}. On the other hand, itinerant approaches
(at weak and strong coupling) have found parameter regimes in which
$g$ and $w$ can be either positive or negative \cite{Eremin10,Fernandes12,xiaoyu14,Wang_arxiv_14,Kang15,Lorenzana08,giovannetti,Brydon11,Berg10}.
For $g>\max\left(0,-w\right)$, the energy is minimized by the stripe
state shown in Fig. \ref{fig_magnetic_gs}a, in which either $\left\langle \mathbf{M}_{1}\right\rangle =0$
or $\left\langle \mathbf{M}_{2}\right\rangle =0$. Thus, in addition
to breaking the $O(3)$ spin-rotational symmetry, the magnetic ground
state spontaneously breaks a $Z_{2}$ symmetry by selecting one of
the two order parameters to be non-zero. Since $\mathbf{M}_{1}$ and
$\mathbf{M}_{2}$ are related by a $90^{\circ}$ rotation, once this
$Z_{2}$ symmetry is broken the tetragonal symmetry of the system
is lowered to orthorhombic (see Fig. \ref{fig_emergent_states}a).
A composite Ising-nematic order parameter, living on the bonds of
the lattice, can be identified by performing a Hubbard-Stratonovich
transformation on the quartic term with coefficient $g$, yielding
$\left\langle \varphi_{\mathrm{nem}}\right\rangle =g\left\langle \mathbf{M}_{1}^{2}-\mathbf{M}_{2}^{2}\right\rangle $.
Because $Z_{2}$ is a discrete symmetry, while spin-rotational $O(3)$
is a continuous symmetry, a strongly anisotropic 3D system will generically
display a vestigial paramagnetic nematic phase where $\left\langle \mathbf{M}_{i}\right\rangle =0$
but $\left\langle \varphi_{\mathrm{nem}}\right\rangle \neq0$ \cite{Fang08,Fernandes12,Batista11}.

For $g<\max\left(0,-w\right)$, the ground state of Eq. (\ref{action})
is no longer a uniaxial magnetic stripe state, but a biaxial magnetic
state with $|\left\langle \mathbf{M}_{1}\right\rangle |=|\left\langle \mathbf{M}_{2}\right\rangle |$
that preserve tetragonal symmetry. If $w<0$, the energy is minimized
by $\left\langle \mathbf{M}_{1}\right\rangle \parallel\left\langle \mathbf{M}_{2}\right\rangle $,
which in terms of the local spin configuration $\mathbf{S}\left(\mathbf{r}\right)$
corresponds to a non-uniform state as depicted in Fig. \ref{fig_magnetic_gs}b.
We identify this state as a charge-spin density-wave (CSDW). On the
other hand, if $w>0$, the energy minimization gives $\left\langle \mathbf{M}_{1}\right\rangle \perp\left\langle \mathbf{M}_{2}\right\rangle $,
corresponding to a non-collinear, coplanar spin configuration (see
Fig. \ref{fig_magnetic_gs}c). This state is identified as a spin
vortex-crystal (SVC). We now discuss whether these tetragonal magnetic
phases can melt in a two-stage process, giving rise to vestigial orders
akin to the nematic phase.

\subsection{Charge-spin density-wave}

Consider the CSDW state: Once the magnetization direction is chosen
by spontaneous breaking of the $O(3)$ spin-rotational symmetry, there
remains a four-fold degeneracy corresponding to whether $\mathbf{M}_{1}$
and $\mathbf{M}_{2}$ are parallel or anti-parallel to the chosen
direction. As is apparent in Fig. \ref{fig_magnetic_gs}b, this corresponds
to the breaking of translational symmetry, leading to a four-site
unit cell. Notice, however, that the product of a translation by the
vector $\hat{\mathbf{x}}+\hat{\mathbf{y}}$ followed by time-reversal
is preserved. Thus, there is an essential $Z_{2}$ symmetry that interchanges
the magnetic and non-magnetic sublattices of the CSDW state.

The order parameter field for this $Z_{2}$ symmetry is obtained via
a Hubbard-Stratonovich transformation on the quartic term with coefficient
$2w$ in Eq. (\ref{action}), $\left\langle \varphi_{\mathrm{CDW}}\right\rangle =2\left|w\right|\left\langle \mathbf{M}_{1}\cdot\mathbf{M}_{2}\right\rangle $.
Clearly, $\varphi_{\mathrm{CDW}}$ is a scalar that carries momentum
$\mathbf{Q}_{1}+\mathbf{Q}_{2}=\left(\pi,\pi\right)$, i.e. the condensed
phase is a CDW that doubles the unit cell, but leaves time-reversal
and the tetragonal symmetry of the lattice intact (see Fig. \ref{fig_emergent_states}b).
Thus, in real space, the CDW order parameter lives on the lattice
sites. The fact that the unit cell decreases from four to two sites
upon going from the CSDW to the CDW phase is due to the restoration
of time-reversal symmetry, which implies the restoration of the translational
symmetry by $\hat{\mathbf{x}}+\hat{\mathbf{y}}$. A simple change
of variables in Eq. (\ref{action}), $\mathbf{M}_{1}\to2^{-1/2}[\mathbf{M}_{1}+\mathbf{M}_{2}]$
and $\mathbf{M}_{1}\to2^{-1/2}[\mathbf{M}_{1}-\mathbf{M}_{2}]$, interchanges
the identities of the two scalar orders, $\varphi_{\mathrm{nem}}\leftrightarrow\varphi_{\mathrm{CDW}}$,
but leaves the form of $\mathcal{S}$ unchanged albeit with $\left(g,w\right)\rightarrow-\left(w,g\right)$.
Thus, the properties of the CDW phase are akin to those of the Ising-nematic
phase -- in particular, a quasi-2D system will again display for a
range of intermediate temperatures a phase with $\left\langle \mathbf{M}_{i}\right\rangle =0$
but $\left\langle \varphi_{\mathrm{CDW}}\right\rangle \neq0$.

\subsection{Spin-vorticity density-wave}

Consider now the SVC state, characterized by two equal magnitude orthogonal
vectors $\mathbf{M}_{1}$ and $\mathbf{M}_{2}$. Upon fixing the direction
of $\mathbf{M}_{1}$, which breaks the $O(3)$ spin-rotational symmetry,
there remains an additional $O(2)$ symmetry related to choosing $\mathbf{M}_{2}$
in any direction along the plane perpendicular to $\mathbf{M}_{1}$
\cite{footnote-symmetries}. Thus, the SVC phase can be completely
characterized by a pseudo-vector order parameter $\boldsymbol{\varphi}_{\mathrm{SVDW}}$
that specifies the ordering plane which contains $\mathbf{M}_{1}$
and $\mathbf{M}_{2}$, and also by the orientation of $\mathbf{M}_{1}$
within that plane. $\boldsymbol{\varphi}_{\mathrm{SVDW}}$ is obtained
via a Hubbard-Stratonovich transformation of the quartic term $w\left(\mathbf{M}_{1}\cdot\mathbf{M}_{2}\right)^{2}\to-w\left(\mathbf{M}_{1}\times\mathbf{M}_{2}\right)^{2}$
in Eq. (\ref{action}), yielding $\left\langle \boldsymbol{\varphi}_{\mathrm{SVDW}}\right\rangle =2w\left\langle \mathbf{M}_{1}\times\mathbf{M}_{2}\right\rangle $,
which can be identified as a vector chirality \cite{Batista09,footnote-inversion}.
Thus, upon approaching the SVC phase from high-temperatures or by
melting it, there can be an intermediate state where $\left\langle \boldsymbol{\varphi}_{\mathrm{SVDW}}\right\rangle \neq0$
but the orientation of $\mathbf{M}_{1}$ is not fixed, $\left\langle \mathbf{M}_{1}\right\rangle =0$.
This chiral paramagnetic state preserves time-reversal symmetry and
retains the memory of the staggering pattern of spin vortices along
the plaquettes in the SVC phase, and is therefore called a spin-vorticity
density-wave (SVDW) \cite{footnote-inversion}. Note that the vector
chiral order parameter produces an emergent Dzyaloshinskii-Moriya
coupling $\boldsymbol{\varphi}_{\mathrm{SVDW}}\cdot\left(\mathbf{M}_{1}\times\mathbf{M}_{2}\right)$
relating the translational symmetry-breaking to a preferred \textquotedblleft handedness\textquotedblright{}
in spin-space. In the SVDW state, not only is the translational symmetry
lowered by the doubling of the unit cell (since $\boldsymbol{\varphi}_{\mathrm{SVDW}}$
carries momentum $\mathbf{Q}_{1}+\mathbf{Q}_{2}=\left(\pi,\pi\right)$),
but also the soft spin fluctuations near the magnetic transition are
constrained to lie in the plane defined by $\boldsymbol{\varphi}_{\mathrm{SVDW}}$
(see Fig. \ref{fig_emergent_states}c).

\begin{figure}
\begin{centering}
\includegraphics[width=0.7\columnwidth]{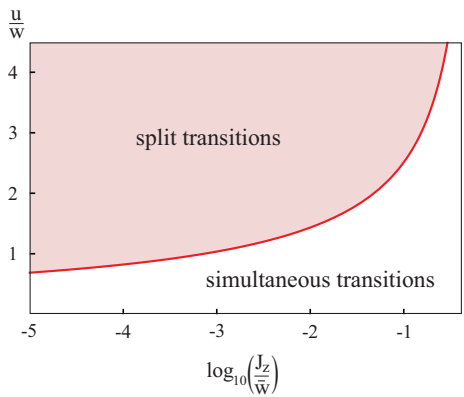} 
\par\end{centering}

\protect\protect\protect\protect\protect\caption{Phase diagram, within the saddle-point approximation, of the coupled
SVDW paramagnetic and SVC magnetic transitions. In the shaded area,
where the out-of-plane anisotropy is strong, the two transitions are
split. The tuning parameters are the Ginzburg-Landau coefficients
$u$ and $w$ (see Eq. \ref{action}) and the magnetic coupling between
layers $J_{z}$. The parameter $\bar{w}$ is given by $w=T_{N,0}w/2\pi$,
as discussed in Appendix \ref{Appendix_free_energy}. \label{fig_phase_diagram}}
\end{figure}

Because $\boldsymbol{\varphi}_{\mathrm{SVDW}}$ breaks a continuous
$O(3)$ symmetry, there are two Goldstone modes in the SVDW phase.
Consequently, in contrast to the Ising-nematic cases, the Mermin-Wagner
theorem does not ensure the existence of the SVDW phase even in the
two-dimensional limit. To investigate whether $\left\langle \boldsymbol{\varphi}_{\mathrm{SVDW}}\right\rangle \neq0$
while $\left\langle \mathbf{M}_{1}\right\rangle =0$ is possible,
we calculated the phase diagram for a magnetic SVC ground state treating
the action in Eq.~(\ref{action}) in the saddle point approximation
(see Appendix \ref{Appendix_saddle_point}). We find that for a strongly
anisotropic system, i.e. $J_{z}\ll w$, there is a wide range of values
of $u/w$ for which there are two transitions, with an intermediate
SVDW phase and a low-temperature SVC phase (see Fig. \ref{fig_phase_diagram}).
However, in this approximation, the transition to the SVDW phase is
always first-order.

Spin rotational symmetry is not an exact symmetry of nature, and indeed
most iron pnictides display a sizable spin anisotropy \cite{Matan09,Dai13,Tucker14}.
Because the ordered moments tend to point parallel to the FeAs plane,
the most significant effects of spin-orbit coupling can be captured
phenomenologically in Eq. (\ref{action}) by including an easy-plane
anisotropy term $\kappa\left(M_{1,z}^{2}+M_{2,z}^{2}\right)$ with
coupling constant $\kappa>0$ \cite{Christensen15}. The spin rotational
symmetry is thus reduced to $O(2)$ and the SVDW chiral order parameter
becomes the pseudo-scalar $\varphi_{\mathrm{SVDW}}=2w\left(\mathbf{M}_{1}\times\mathbf{M}_{2}\right)\cdot\hat{\mathbf{z}}$,
which only breaks a discrete chiral $Z_{2}$ symmetry. For such $O(2)\times Z_{2}$
model, it is known from both numerical and analytical investigations
that in $2$D the $Z_{2}$ symmetry is broken at higher temperatures
than the Kosterlitz-Thouless transition of the $O(2)$ order parameter
\cite{Korshunov02,Vicari05}, \textit{i.e.} there is no doubt that
there is a vestigial chiral SVDW phase. The extent to which the spin
anisotropy is quantitatively significant depends on the (currently
unknown) value of the ratio $\kappa/\left(T_{\mathrm{\mathrm{SVDW}}}-T_{\mathrm{SVC}}\right)$.

\section{Microscopic implications of the vestigial orders \label{sec_Microscopics}}

\subsection{Normal-state manifestations}

\begin{figure}
\begin{centering}
\includegraphics[width=0.75\columnwidth]{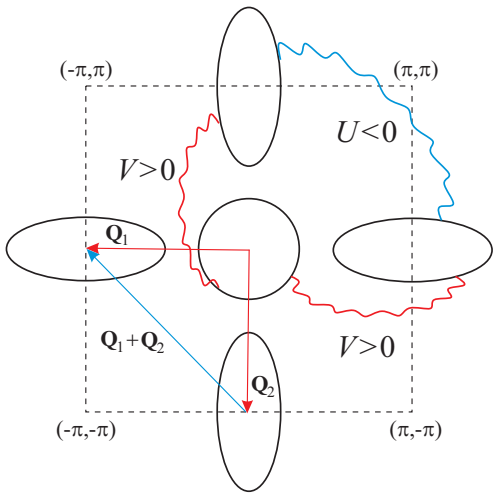} 
\par\end{centering}

\protect\protect\protect\protect\protect\caption{Schematic Fermi surface of the iron pnictides, with a central hole
pocket and elliptical electron pockets. The wavy lines represent the
inter-pocket pairing interactions generated by the magnetic fluctuations
(repulsive $V>0$) and by fluctuations of the vestigial CDW and SVDW
states (attractive $U<0$). \label{fig_superconductivity}}
\end{figure}

To discuss the experimental manifestations of the vestigial CDW and
SVDW states, we investigate their coupling to the low-energy electronic
states of the pnictides. We consider a three-band model \cite{Eremin10,Fernandes12}
with a circular hole pocket $\xi_{h,\mathbf{k}}$ at the center of
the Brillouin zone, and two elliptical electron pockets $\xi_{e_{1},\mathbf{k}+\mathbf{Q}_{1}}$
and $\xi_{e_{2},\mathbf{k}+\mathbf{Q}_{2}}$ centered at momenta $\mathbf{Q}_{1}=\left(\pi,0\right)$
and $\mathbf{Q}_{2}=\left(0,\pi\right)$, respectively (see Fig. \ref{fig_superconductivity}).
The magnetic order parameters couple to these electronic states via
$\sum_{\mathbf{k},\alpha\beta}\mathbf{M}_{i}\cdot\boldsymbol{\sigma}_{\alpha\beta}\left(c_{h,\mathbf{k}\alpha}^{\dagger}c_{e_{i},\mathbf{k}\beta}^{\phantom{\dagger}}+\mathrm{h.c.}\right)$,
where the operator $c_{a,\mathbf{k}\alpha}$ annihilates an electron
in band $a$ with momentum $\mathbf{k}$ (measured with respect to
the center of the pocket) and spin $\alpha$, and $\boldsymbol{\sigma}_{\alpha\beta}$
are Pauli matrices. We further introduce magnetic $\boldsymbol{\Delta}_{S}$
and charge $\Delta_{C}$ order parameters with ordering vector $\mathbf{Q}_{1}+\mathbf{Q}_{2}=\left(\pi,\pi\right)$,
which couple to the electronic states via

\begin{table*}[t]
\protect\protect\protect\protect\protect\caption{Magnetic ground states of the pnictides and their corresponding vestigial
states. $\mathbf{M}_{1}$ and $\mathbf{M}_{2}$ are the magnetic order
parameters corresponding to the ordering vectors $\mathbf{Q}_{1}=\left(\pi,0\right)$
and $\mathbf{Q}_{2}=\left(0,\pi\right)$. \label{table_summary}}

\centering{}%
\begin{tabular}{|l|l|l|l|l|}
\hline 
\textbf{\emph{magnetic ground state}}  & \textbf{\emph{vestigial state}}  & \textbf{\emph{broken symmetry}}  & \textbf{\emph{real space pattern}}  & \textbf{\emph{physical manifestation }}\tabularnewline
\hline 
\hline 
stripe: $\left\langle \mathbf{M}_{2}\right\rangle \mathrm{\, or}\,\left\langle \mathbf{M}_{1}\right\rangle =0$  & nematic: $\left\langle \mathbf{M}_{1}^{2}-\mathbf{M}_{2}^{2}\right\rangle \neq0$  & rotational (tetragonal)  & unequal bonds  & orthorhombic distortion\tabularnewline
\hline 
CSDW: $\left\langle \mathbf{M}_{1}\right\rangle \parallel\left\langle \mathbf{M}_{2}\right\rangle $  & CDW: $\left\langle \mathbf{M}_{1}\cdot\mathbf{M}_{2}\right\rangle \neq0$  & translational  & unequal sites  & charge density-wave\tabularnewline
\hline 
SVC: $\left\langle \mathbf{M}_{1}\right\rangle \perp\left\langle \mathbf{M}_{2}\right\rangle $  & SVDW: $\left\langle \mathbf{M}_{1}\times\mathbf{M}_{2}\right\rangle \neq\mathbf{0}$  & translational + inversion  & unequal plaquettes  & spin-current density-wave \tabularnewline
\hline 
\end{tabular}
\end{table*}
\begin{align}
\mathcal{H}_{S} & =\sum_{\mathbf{k},\alpha\beta}\left[\boldsymbol{\Delta}_{S}
\cdot\boldsymbol{\sigma}_{\alpha\beta}^{\phantom{\dagger}}c_{e_{2},\mathbf{k}\alpha}^{\dagger}c_{e_{1},\mathbf{k}\beta}^{\phantom{\dagger}}+\mathrm{h.c.}\right],\nonumber \\
\mathcal{H}_{C} & =\sum_{\mathbf{k},\alpha\beta}\left[\Delta_{C}
\delta_{\alpha\beta}^{\phantom{\dagger}}c_{e_{2},\mathbf{k}\alpha}^{\dagger}c_{e_{1},\mathbf{k}\beta}^{\phantom{\dagger}}+\mathrm{h.c.}\right]\label{eq_CDW_SDW-2}
\end{align}

Here these fields have real and imaginary parts, $\boldsymbol{\Delta}_{S}=\boldsymbol{\Delta}_{S}^{\prime}+i\boldsymbol{\Delta}_{S}^{\prime\prime}$
and $\Delta_{C}=\Delta_{C}^{\prime}+i\Delta_{C}^{\prime\prime}$,
where the real parts correspond to conventional SDW or CDW orders,
while the imaginary parts corresponds to spin or charge current orders.
By integrating out the electronic degrees of freedom, we obtain the
coupling between $\mathbf{M}_{i}$ and $\boldsymbol{\Delta}_{S}
$, $\Delta_{C}
$ to lowest-order in the action (see Appendix \ref{Appendix_free_energy}):

\begin{equation}
\delta\mathcal{S}_{\mathrm{eff}}=\lambda\left[\boldsymbol{\Delta}_{S}^{\prime\prime}\cdot\left(\mathbf{M}_{1}\times\mathbf{M}_{2}\right)-\Delta_{C}^{\prime}\left(\mathbf{M}_{1}\cdot\mathbf{M}_{2}\right)\right]\label{eq_S_Eff}
\end{equation}
with the coefficient $\lambda=4\int_{k}G_{h,k}G_{e_{1},k}G_{e_{2},k}$,
where $G_{a,k}^{-1}=i\omega_{n}-\xi_{a,\mathbf{k}}$ is the corresponding
non-interacting Green's function. As expected, the Ising-like order
parameter $\varphi_{\mathrm{CDW}}\propto\mathbf{M}_{1}\cdot\mathbf{M}_{2}$
induces a checkerboard-like charge order (see Fig. \ref{fig_emergent_states}b).
On the other hand, the SVDW order parameter $\boldsymbol{\varphi}_{\mathrm{SVDW}}\propto\mathbf{M}_{1}\times\mathbf{M}_{2}$
is manifested as a spin-current density-wave with propagation vector
$\left(\pi,\pi\right)$, i.e. a spin current polarized parallel to
$\boldsymbol{\varphi}_{\mathrm{SVDW}}$ and propagating along the
bonds of the lattice in a staggered pattern across the square plaquettes
(see Fig. \ref{fig_emergent_states}c). Thus, the SVDW corresponds
to a triplet $d$-density wave \cite{Chakravarty01}.

Note that probing the CDW via x-rays may be difficult, since the hybridization
between Fe and As/Se doubles the unit cell of the Fe-only square lattice,
making $\left(\pi,\pi\right)$ a lattice Bragg peak. While the real
CDW could in principle be detected experimentally by a probe sensitive
to the local charge on the Fe sites, such as STM, detecting a spin-current
density-wave would be rather challenging. Alternatively, one can consider
the effects of a Zeeman field $\mathbf{H}$. Despite not coupling
to $\varphi_{\mathrm{nem}}$, we find that it couples to both $\varphi_{\mathrm{CDW}}$
and $\boldsymbol{\varphi}_{\mathrm{SVDW}}$ in the action via the
terms $\gamma\Delta_{C}^{\prime\prime}
\mathbf{H}\cdot\left(\mathbf{M}_{1}\times\mathbf{M}_{2}\right)$ and $\gamma\left(\mathbf{H}\cdot\boldsymbol{\Delta}_{S}^{\prime}
\right)\left(\mathbf{M}_{1}\cdot\mathbf{M}_{2}\right)$, with the same Ginzburg-Landau coefficient $\gamma$. Therefore,
in the presence of a magnetic field, a pattern of staggering orbital
currents (i.e. a singlet $d$-density wave \cite{Chakravarty01})
appears in the SVDW state, which can in principle be detected by NMR.
Table \ref{table_summary} summarizes the magnetic ground states of
the pnictides along their vestigial paramagnetic states.

\subsection{Impact on the superconducting state}

Fluctuations of the SVDW and CDW states arise from four-spin correlations,
and are complementary to the magnetic fluctuations that arise from
two-spin correlations. An important issue is whether these fluctuation
modes promote compatible superconducting states.\textcolor{red}{{} }Because
the magnetic fluctuations are peaked at momenta $\mathbf{Q}_{1}=\left(\pi,0\right)$
and $\mathbf{Q}_{2}=\left(0,\pi\right)$, they promote a \emph{repulsive}
inter-pocket interaction $V>0$ between the hole and the electron
pockets (see Fig. \ref{fig_superconductivity}). Solution of the corresponding
linearized gap equations yields the so-called $s^{+-}$ state, where
the gap functions have different signs in the electron and in the
hole pockets \cite{reviews_pairing}. The transition temperature is
given by $T_{c}\propto\exp\left(-\frac{1}{\Lambda_{0}}\right)$, with
$\Lambda_{0}=\sqrt{2N_{h}N_{e}}\ V$, and $N_{a}$ denoting the density
of states of band $a$. 

The SVDW and CDW fluctuations, on the other hand, are peaked at the
momentum $\mathbf{Q}_{1}+\mathbf{Q}_{2}=\left(\pi,\pi\right)$ and
promote an \emph{attractive} inter-pocket interaction $U<0$ between
the two electron pockets (see Fig. \ref{fig_superconductivity} and
Appendix \ref{Appendix_pairing}). Solution of the linearized gap
equation reveals that the leading eigenstate remains the $s^{+-}$
one, but the eigenvalue is enhanced, $\Lambda=\sqrt{\Lambda_{0}^{2}+\Lambda_{U}^{2}}+\Lambda_{U}$,
with $\Lambda_{U}=\frac{N_{e}\left|U\right|}{2}>0$. Therefore, fluctuations
associated with these vestigial states may enhance the value of $T_{c}$
promoted by spin-fluctuations pairing, without affecting the symmetry
of the Cooper pair wave-function. Similar conclusions have been found
for the combination of pairing promoted by nematic fluctuations (peaked
at $\mathbf{Q}=0$) and magnetic fluctuations (peaked at $\mathbf{Q}_{1}$
and $\mathbf{Q}_{2}$) \cite{Lederer14}.

\section{Concluding remarks \label{sec_Concluding_remarks}}

In summary, we showed that both biaxial tetragonal magnetic ground
states of the pnictides -- the non-uniform CSDW and non-collinear
SVC states -- can melt in two-stage processes, giving rise to CDW
and SVDW vestigial states, respectively. While both preserve the point-group
and time-reversal symmetries, but break the translational symmetry
of the iron square lattice, only the SVDW state also breaks inversion
symmetry by entangling the spin-space handedness to a doubling of
the real-space unit cell. Because in the iron superconductors the
hybridization with the puckered As atoms already doubles the unit
cell of the Fe square lattice, the CDW and SVDW states are more rigorously
classified as intra-unit-cell orders. Recent experiments on $\mathrm{Sr_{1-x}Na_{x}Fe_{2}As_{2}}$
\cite{Allred15} and $\mathrm{Ba_{1-x}K_{x}Fe_{2}As_{2}}$ \cite{Mallet15}
found direct evidence for a low-temperature CSDW phase, which can
support a CDW vestigial phase. It remains to be seen whether the tetragonal
magnetic phase can be reached in these compounds without first crossing
the stripe magnetic state. In contrast, in $\text{Ba}(\text{Fe}_{1-x}\text{Mn}_{x})_{2}\text{As}_{2}$
\cite{Kim10}, the tetragonal magnetic phase has been reported to
exist over a wide doping range as the primary instability of the paramagnetic
phase.

Beyond the physics of iron-based superconductors, our results establish
the melting of double-\textbf{Q} orders as a microscopic mechanism
to create $d$-density wave states. The latter have been proposed
to be realized in other strongly correlated systems, such as the pseudogap
phase of underdoped cuprates \cite{Chakravarty01} and the hidden-order
phase of the heavy fermion compound $\mathrm{URu_{2}Si_{2}}$ \cite{Fujimoto11,Chakravarty14,Blumberg15},
mostly on phenomenological basis. Whether our mechanism is directly
applicable to those systems is an appealing topic for future investigation.
\begin{acknowledgments}
We thank C. Batista, A. Boehmer, A. Chubukov, J. Kang, C. Meingast,
R. Osborn, M. Schuett, and X. Wang for fruitful discussions. RMF is
supported by the U.S. Department of Energy, Office of Science, Basic
Energy Sciences, under award number DE-SC0012336. SAK is supported
by the U.S. Department of Energy under Contract No. DE-AC02-76SF00515.
EB was supported by the Israeli Science Foundation, by the US-Israel
Binational Science Foundation, and by an Alon fellowship. We thank
the hospitality of the Aspen Center for Physics, where this work was
initiated. \end{acknowledgments}

\appendix
\begin{widetext}

\section{Saddle-point equations for the SVDW order \label{Appendix_saddle_point}}

We start with the effective action for the magnetic order parameters:

\begin{equation}
S\left[\mathbf{M}_{i}\right]=\int_{q}\chi_{q}^{-1}\left(M_{1}^{2}+M_{2}^{2}\right)+\frac{u}{2}\int_{x}\left(M_{1}^{2}+M_{2}^{2}\right)^{2}-\frac{g}{2}\int_{x}\left(M_{1}^{2}-M_{2}^{2}\right)^{2}+2w\int_{x}\left(\mathbf{M}_{1}\cdot\mathbf{M}_{2}\right)^{2}\label{F_mag}
\end{equation}
where $\int_{q}\equiv T\sum_{\omega_{n}}\int\frac{d^{d}q}{\left(2\pi\right)^{d}}$
and $\int_{x}\equiv\int_{0}^{\beta}d\tau\int d^{d}x$. To proceed,
we use the identity:

\begin{equation}
\left(\mathbf{M}_{1}\cdot\mathbf{M}_{2}\right)^{2}=\frac{1}{4}\left(M_{1}^{2}+M_{2}^{2}\right)^{2}-\frac{1}{4}\left(M_{1}^{2}-M_{2}^{2}\right)^{2}-\left(\mathbf{M}_{1}\times\mathbf{M}_{2}\right)^{2}
\end{equation}
yielding:

\begin{equation}
S\left[\mathbf{M}_{i}\right]=\int_{q}\chi_{q}^{-1}\left(M_{1}^{2}+M_{2}^{2}\right)+\frac{\left(u+w\right)}{2}\int_{x}\left(M_{1}^{2}+M_{2}^{2}\right)^{2}-\frac{\left(g+w\right)}{2}\int_{x}\left(M_{1}^{2}-M_{2}^{2}\right)^{2}-2w\int_{x}\left(\mathbf{M}_{1}\times\mathbf{M}_{2}\right)^{2}
\end{equation}

Hereafter for simplicity we introduce the parameters $\tilde{g}=g+w$
and $\tilde{u}=u+w$. Since we are interested in the vestigial phase
of the spin vortex-crystal, which has tetragonal symmetry, the nematic
order parameter $\left\langle M_{1}^{2}\right\rangle -\left\langle M_{2}^{2}\right\rangle $
never condenses, and we can ignore the corresponding quartic term.
Introducing the Hubbard-Stratonovich fields corresponding to the other
two quadratic terms, we obtain:

\begin{eqnarray}
\mathrm{e}^{-\frac{u}{2}\left(M_{1}^{2}+M_{2}^{2}\right)^{2}} & = & \mathcal{N}\int d\psi\,\mathrm{e}^{\frac{\psi^{2}}{2u}-\psi\left(M_{1}^{2}+M_{2}^{2}\right)}\nonumber \\
\mathrm{e}^{2w\left(\mathbf{M}_{1}\times\mathbf{M}_{2}\right)\cdot\left(\mathbf{M}_{1}\times\mathbf{M}_{2}\right)} & = & \mathcal{N}\int d\boldsymbol{\varphi}_{\mathrm{SVDW}}\,\mathrm{e}^{\frac{-\varphi^{2}}{2w}+2\boldsymbol{\varphi}_{\mathrm{SVDW}}\cdot\left(\mathbf{M}_{1}\times\mathbf{M}_{2}\right)}
\end{eqnarray}

Here, $\boldsymbol{\varphi}_{\mathrm{SVDW}}$ is the spin-vorticity
density-wave (SVDW) vectorial order parameter whose mean value is
given by $\left\langle \boldsymbol{\varphi}_{\mathrm{SVDW}}\right\rangle =2w\left\langle \mathbf{M}_{1}\times\mathbf{M}_{2}\right\rangle $.
The field $\psi$ is not an order parameter, and just renormalizes
the magnetic correlation length via $\left\langle \psi\right\rangle =\tilde{u}\left(\left\langle M_{1}^{2}\right\rangle +\left\langle M_{2}^{2}\right\rangle \right)$,
i.e. it corresponds to Gaussian magnetic fluctuations. Thus, the effective
action is given by:

\begin{equation}
S\left[\mathbf{M}_{i},\psi,\boldsymbol{\varphi}_{\mathrm{SVDW}}\right]=\int_{q}\left(\chi_{q}^{-1}+\psi\right)\left(M_{1}^{2}+M_{2}^{2}\right)-2\int_{x}\boldsymbol{\varphi}_{\mathrm{SVDW}}\cdot\left(\mathbf{M}_{1}\times\mathbf{M}_{2}\right)+\frac{\varphi_{\mathrm{SVDW}}^{2}}{2w}-\frac{\psi^{2}}{2\tilde{u}}
\end{equation}

Approaching the SVDW phase from the paramagnetic state, we can integrate
out the magnetic degrees of freedom, yielding an effective action
for $\psi$ and $\boldsymbol{\varphi}_{\mathrm{SVDW}}$:

\begin{equation}
S_{\mathrm{eff}}\left[\psi,\boldsymbol{\varphi}_{\mathrm{SVDW}}\right]=\frac{\varphi_{\mathrm{SVDW}}^{2}}{2w}-\frac{\psi^{2}}{2\tilde{u}}+\frac{1}{2}\int_{q}\log\left(\prod_{i}\lambda_{i,q}\right)
\end{equation}
where $\lambda_{i,q}$ are the eigenvalues of the matrix $A_{ij}$
corresponding to the Gaussian action in $\mathbf{M}_{i}$. The Gaussian
part of the action can be rewritten in the convenient matrix form:

\begin{equation}
\left(\begin{array}{cc}
\mathbf{M}_{1} & \mathbf{M}_{2}\end{array}\right)\left(\begin{array}{cccccc}
\chi_{q}^{-1}+\psi & 0 & 0 & 0 & -\varphi_{z} & \varphi_{y}\\
0 & \chi_{q}^{-1}+\psi & 0 & \varphi_{z} & 0 & -\varphi_{x}\\
0 & 0 & \chi_{q}^{-1}+\psi & -\varphi_{y} & \varphi_{x} & 0\\
0 & \varphi_{z} & -\varphi_{y} & \chi_{q}^{-1}+\psi & 0 & 0\\
-\varphi_{z} & 0 & \varphi_{x} & 0 & \chi_{q}^{-1}+\psi & 0\\
\varphi_{y} & -\varphi_{x} & 0 & 0 & 0 & \chi_{q}^{-1}+\psi
\end{array}\right)\left(\begin{array}{c}
\mathbf{M}_{1}\\
\mathbf{M}_{2}
\end{array}\right)\label{eq_matrix_mag}
\end{equation}

Evaluation of the eigenvalues gives:

\begin{equation}
S_{\mathrm{eff}}\left[\psi,\boldsymbol{\varphi}_{\mathrm{SVDW}}\right]=\frac{\varphi_{\mathrm{SVDW}}^{2}}{2w}-\frac{\psi^{2}}{2\tilde{u}}+\int_{q}\log\left[\left(\chi_{q}^{-1}+\psi\right)\left(\chi_{q}^{-1}+\psi+\varphi_{\mathrm{SVDW}}\right)\left(\chi_{q}^{-1}+\psi-\varphi_{\mathrm{SVDW}}\right)\right]\label{S_eff_final}
\end{equation}

So far our result is exact. To proceed, we employ the saddle-point
approximation to determine the equations of state for $\psi$ and
$\varphi_{\mathrm{SVDW}}$, which corresponds to self-consistently
accounting for the Gaussian magnetic fluctuations. The saddle-point
equations become:

\begin{eqnarray}
\frac{\varphi_{\mathrm{SVDW}}}{w} & = & \int_{q}\frac{1}{\chi_{q}^{-1}+\psi-\varphi_{\mathrm{SVDW}}}-\int_{q}\frac{1}{\chi_{q}^{-1}+\psi+\varphi_{\mathrm{SVDW}}}\nonumber \\
\frac{\psi}{\tilde{u}} & = & \int_{q}\frac{1}{\chi_{q}^{-1}+\psi-\varphi_{\mathrm{SVDW}}}+\int_{q}\frac{1}{\chi_{q}^{-1}+\psi+\varphi_{\mathrm{SVDW}}}+\int_{q}\frac{1}{\chi_{q}^{-1}+\psi}
\end{eqnarray}

Since our focus is on the proximity to a finite-temperature magnetic
transition, we ignore the spin dynamics and use the low-energy expansion
for the spin susceptibility appropriate for anisotropic layered systems:

\begin{equation}
\chi_{q}^{-1}=r_{0}+q_{\parallel}^{2}+J_{z}\sin^{2}\frac{q_{z}}{2}
\end{equation}
where $r_{0}=a\left(T-T_{N}\right)$, $a>0$, $T_{N}$ is the mean-field
magnetic transition temperature, $q_{\parallel}^{2}=q_{x}^{2}+q_{y}^{2}$,
and $J_{z}$ is the inter-layer magnetic coupling. Defining the renormalized
magnetic mass:

\begin{equation}
r=r_{0}+\psi\propto\xi^{-2}
\end{equation}
where $\xi$ is the magnetic correlation length, we obtain:

\begin{eqnarray}
\varphi_{\mathrm{SVDW}} & = & w\left[\int_{q}\frac{1}{r+q_{\parallel}^{2}+J_{z}\sin^{2}\frac{q_{z}}{2}-\varphi_{\mathrm{SVDW}}}-\int_{q}\frac{1}{r+q_{\parallel}^{2}+J_{z}\sin^{2}\frac{q_{z}}{2}+\varphi_{\mathrm{SVDW}}}\right]\\
r & = & r_{0}+\tilde{u}\left[\int_{q}\frac{1}{r+q_{\parallel}^{2}+J_{z}\sin^{2}\frac{q_{z}}{2}-\varphi_{\mathrm{SVDW}}}+\int_{q}\frac{1}{r+q_{\parallel}^{2}+J_{z}\sin^{2}\frac{q_{z}}{2}+\varphi_{\mathrm{SVDW}}}+\int_{q}\frac{1}{r+q_{\parallel}^{2}+J_{z}\sin^{2}\frac{q_{z}}{2}}\right]\nonumber 
\end{eqnarray}

The integrals can be evaluated in a straightforward way (we consider
only the $\omega_{n}=0$ contribution to the sum over Matsubara frequencies,
since we are interested in the finite temperature transition):

\begin{eqnarray}
\int_{q}\frac{1}{q_{\parallel}^{2}+J_{z}\sin^{2}\frac{q_{z}}{2}+a} & = & \frac{T_{N}}{4\pi}\int_{0}^{2\pi}\frac{dq_{z}}{2\pi}\int_{J_{z}\sin^{2}\frac{q_{z}}{2}+a}^{\Lambda^{2}}\frac{dx}{x}\nonumber \\
 & = & \frac{T_{N}}{4\pi}\int_{0}^{2\pi}\frac{dq_{z}}{2\pi}\ln\left(\frac{\Lambda^{2}}{J_{z}\sin^{2}\frac{q_{z}}{2}+a}\right)\nonumber \\
 & = & \frac{T_{N}}{2\pi}\left[\ln2\Lambda-\ln\left(\sqrt{J_{z}+a}+\sqrt{a}\right)\right]
\end{eqnarray}

Defining the renormalized critical temperature $\tilde{r}_{0}=a(T-\tilde{T}_{N})$
via:

\begin{equation}
\tilde{r}_{0}=r_{0}+\frac{3\tilde{u}T_{N}}{2\pi}\ln\frac{2\Lambda}{\sqrt{J_{z}}}
\end{equation}
we obtain the self-consistent equations:

\begin{eqnarray}
\varphi_{\mathrm{SVDW}} & = & \frac{wT_{N}}{2\pi}\ln\frac{\sqrt{J_{z}+r+\varphi_{\mathrm{SVDW}}}+\sqrt{r+\varphi_{\mathrm{SVDW}}}}{\sqrt{J_{z}+r-\varphi_{\mathrm{SVDW}}}+\sqrt{r-\varphi_{\mathrm{SVDW}}}}\\
r & = & \tilde{r}_{0}-\frac{\tilde{u}T_{N}}{2\pi}\ln\left[\frac{\left(\sqrt{J_{z}+r+\varphi_{\mathrm{SVDW}}}+\sqrt{r+\varphi_{\mathrm{SVDW}}}\right)\left(\sqrt{J_{z}+r-\varphi_{\mathrm{SVDW}}}+\sqrt{r-\varphi_{\mathrm{SVDW}}}\right)\left(\sqrt{J_{z}+r}+\sqrt{r}\right)}{J_{z}^{3/2}}\right]\nonumber 
\end{eqnarray}

For simplicity, we define the renormalized parameters $\left(\bar{w},\bar{u}\right)\equiv\left(w,u\right)\frac{T_{N}}{2\pi}$
as well as $\alpha\equiv\frac{\tilde{u}}{w}=\frac{u}{w}+1$ and $\tilde{J}_{z}\equiv J_{z}/\bar{w}$.
Then the equations can be written as:

\begin{eqnarray}
\varphi_{\mathrm{SVDW}} & = & \ln\frac{\sqrt{\tilde{J}_{z}+r+\varphi_{\mathrm{SVDW}}}+\sqrt{r+\varphi_{\mathrm{SVDW}}}}{\sqrt{\tilde{J}_{z}+r-\varphi_{\mathrm{SVDW}}}+\sqrt{r-\varphi_{\mathrm{SVDW}}}}\label{eq_saddle_point}\\
r & = & \tilde{r}_{0}-\alpha\ln\left[\frac{\left(\sqrt{\tilde{J}_{z}+r+\varphi_{\mathrm{SVDW}}}+\sqrt{r+\varphi_{\mathrm{SVDW}}}\right)\left(\sqrt{\tilde{J}_{z}+r-\varphi_{\mathrm{SVDW}}}+\sqrt{r-\varphi_{\mathrm{SVDW}}}\right)\left(\sqrt{\tilde{J}_{z}+r}+\sqrt{r}\right)}{\tilde{J}_{z}^{3/2}}\right]\nonumber 
\end{eqnarray}
where $r$, $\tilde{r}_{0}$, and $\varphi_{\mathrm{SVDW}}$ were
rescaled by $\bar{w}$ as well. The SVDW transition temperature can
be obtained by linearizing the equations around $\varphi_{\mathrm{SVDW}}=0$.
From the first equation, we obtain the correlation length $r_{1}$
at the SVDW transition: 

\begin{equation}
r_{1}=\frac{\sqrt{\tilde{J}_{z}^{2}+4}-\tilde{J}_{z}}{2}
\end{equation}
which, when substituted in the second equation, gives the SVDW transition
temperature $\tilde{r}_{0,\mathrm{SVDW}}$:

\begin{equation}
\tilde{r}_{0,\mathrm{SVDW}}=\frac{\sqrt{\tilde{J}_{z}^{2}+4}-\tilde{J}_{z}}{2}+3\alpha\ln\left(\frac{\sqrt{\sqrt{\tilde{J}_{z}^{2}+4}+\tilde{J}_{z}}+\sqrt{\sqrt{\tilde{J}_{z}^{2}+4}-\tilde{J}_{z}}}{\sqrt{2\tilde{J_{z}}}}\right)\label{r0_SVDW}
\end{equation}

The magnetic transition temperature $\tilde{r}_{0,\mathrm{mag}}$
is signaled by the vanishing of the renormalized magnetic mass, i.e.
the lowest eigenvalue of the Eq. (\ref{eq_matrix_mag}), $r-\varphi_{\mathrm{SVDW}}$.
Therefore, it takes place when $r$ reaches the value $r_{2}$ determined
implicitly by:

\begin{equation}
r_{2}=\ln\frac{\sqrt{\tilde{J}_{z}+2r_{2}}+\sqrt{2r_{2}}}{\sqrt{\tilde{J}_{z}}}
\end{equation}

The magnetic transition temperature is therefore given by:

\begin{equation}
\tilde{r}_{0,\mathrm{mag}}=r_{2}\left(1+\alpha\right)+\alpha\ln\left[\frac{\sqrt{\tilde{J}_{z}+r_{2}}+\sqrt{r_{2}}}{\sqrt{\tilde{J}_{z}}}\right]\label{r0_mag}
\end{equation}

The SVDW and magnetic transitions are split when $\tilde{r}_{0,\mathrm{SVDW}}>\tilde{r}_{0,\mathrm{mag}}$.
The region in the $\left(\frac{u}{w},\tilde{J_{z}}\right)$ parameter
space where this condition is satisfied corresponds to the shaded
area of Fig. 3 in the main text (recall that $\frac{u}{w}=\alpha-1$). 

To determine the character of the SVDW transition, we can expand $\tilde{r}_{0}$
for small $\varphi_{\mathrm{SVDW}}$. Substituting $r=r_{1}+a\varphi_{\mathrm{SVDW}}^{2}$
in the first equation of (\ref{eq_saddle_point}) and expanding for
small $\varphi_{\mathrm{SVDW}}$ gives the coefficient of the quadratic
term:

\begin{equation}
a=\frac{8+3\tilde{J}_{z}^{2}}{12\sqrt{\tilde{J}_{z}^{2}+4}}
\end{equation}

Substituting it in the second equation of (\ref{eq_saddle_point})
and collecting the quadratic terms in $\varphi_{\mathrm{SVDW}}$ yields:

\begin{equation}
\tilde{r}_{0}\left(\varphi_{\mathrm{SVDW}}\right)\approx\tilde{r}_{0,\mathrm{SVDW}}+\left[\frac{16+3\tilde{J}_{z}^{2}\left(2+\alpha\right)}{24\sqrt{\tilde{J}_{z}^{2}+4}}\right]\varphi_{\mathrm{SVDW}}^{2}\label{r0_SVDW_exp}
\end{equation}

Therefore, because the coefficient is always positive, the solution
with $\varphi_{\mathrm{SVDW}}\neq0$ is achieved at a larger temperature
than the solution with $\varphi_{\mathrm{SVDW}}=0$, in other words,
$\tilde{r}_{0}\left(\varphi_{\mathrm{SVDW}}>0\right)>\tilde{r}_{0}\left(\varphi_{\mathrm{SVDW}}\rightarrow0\right)$.
As a result, the SVDW transition is first-order within the saddle-point
approximation, even when it is split from the magnetic transition.

\section{Derivation of the Ginzburg-Landau free energy \label{Appendix_free_energy}}

Our starting point is a 3-band model with a circular hole pocket $h$
centered at $\left(0,0\right)$ and two elliptical electron pockets
$e_{1,2}$ centered at $\mathbf{Q}_{1}=\left(\pi,0\right)$ and $\mathbf{Q}_{2}=\left(0,\pi\right)$,
respectively. The band dispersions can be conveniently parametrized
by \cite{Fernandes12}:

\begin{eqnarray}
\xi_{h,\mathbf{k}} & = & -\xi_{\mathbf{k}}=-\frac{k^{2}}{2m}+\varepsilon_{0}\nonumber \\
\xi_{e_{1},\mathbf{k}+\mathbf{Q}_{1}} & = & \xi_{\mathbf{k}}-\left(\delta_{0}+\delta_{2}\cos2\theta\right)\nonumber \\
\xi_{e_{2},\mathbf{k}+\mathbf{Q}_{2}} & = & \xi_{\mathbf{k}}-\left(\delta_{0}-\delta_{2}\cos2\theta\right)\label{band_structure_delta}
\end{eqnarray}

Here, $\delta_{0}$ is proportional to the chemical potential and
$\delta_{2}$ to the ellipticity of the electron pockets. The angle
$\theta$ is measured relative to the $k_{x}$ axis. The non-interacting
Hamiltonian is therefore given by (hereafter sums over repeated spin
indices are implicitly assumed):

\begin{equation}
H_{0}=\sum_{\mathbf{k}}\xi_{h,\mathbf{k}}c_{h,\mathbf{k}\sigma}^{\dagger}c_{h,\mathbf{k}\sigma}+\sum_{\mathbf{k}}\xi_{e_{1},\mathbf{k}}c_{e_{1},\mathbf{k}\sigma}^{\dagger}c_{e_{1},\mathbf{k}\sigma}+\sum_{\mathbf{k}}\xi_{e_{2},\mathbf{k}}c_{e_{2},\mathbf{k}\sigma}^{\dagger}c_{e_{2},\mathbf{k}\sigma}\label{H0}
\end{equation}

These electronic states couple to the magnetic order parameters $\mathbf{M}_{1}$
and $\mathbf{M}_{2}$ according to:

\begin{equation}
H_{\mathrm{mag}}=\sum_{\mathbf{k},i}\mathbf{M}_{i}\cdot\left(c_{e_{i},\mathbf{k}\alpha}^{\dagger}\boldsymbol{\sigma}_{\alpha\beta}c_{h,\mathbf{k}\beta}+\mathrm{h.c.}\right)\label{H_mag}
\end{equation}

In principle, this last term can be obtained via a Hubbard-Stratonovich
transformation of the original interaction terms projected into the
magnetic channel, as shown in Ref. \cite{Fernandes12}. Here, because
we are interested in the higher-order couplings of the action involving
the $\mathbf{M}_{i}$ order parameters, we neglect these interaction
terms, since they only affect the quadratic terms of the action.

\subsection{Absence of magnetic field}

In the case where there is no external magnetic field, we focus on
the two types of fermionic order that couple directly to the SVDW
order parameter, $\mathbf{M}_{1}\times\mathbf{M}_{2}$, and to the
CDW order parameter $\mathbf{M}_{1}\cdot\mathbf{M}_{2}$. Thus, we
introduce the $\mathbf{Q}_{1}+\mathbf{Q}_{2}=\left(\pi,\pi\right)$
spin-current density-wave $\boldsymbol{\Delta}''_{S}$ (i.e. a purely
imaginary SDW) and the checkerboard charge order $\Delta'_{C}$ (i.e.
a purely real CDW) defined by:

\begin{align}
\mathcal{H}_{iS} & =i\sum_{\mathbf{k}}\boldsymbol{\Delta}''_{S}\cdot\boldsymbol{\sigma}_{\alpha\beta}^{\phantom{\dagger}}\left(c_{e_{2},\mathbf{k}\alpha}^{\dagger}c_{e_{1},\mathbf{k}\beta}^{\phantom{\dagger}}-c_{e_{1},\mathbf{k}\alpha}^{\dagger}c_{e_{2},\mathbf{k}\beta}^{\phantom{\dagger}}\right)\nonumber \\
\mathcal{H}_{C} & =\sum_{\mathbf{k}}\Delta'_{C}\delta_{\alpha\beta}\left(c_{e_{2},\mathbf{k}\alpha}^{\dagger}c_{e_{1},\mathbf{k}\beta}^{\phantom{\dagger}}+c_{e_{1},\mathbf{k}\alpha}^{\dagger}c_{e_{2},\mathbf{k}\beta}^{\phantom{\dagger}}\right)\label{eq_CDW_SDW}
\end{align}

To proceed, we introduce the $6$-dimensional Nambu operator:
\begin{equation}
\Psi_{\mathbf{k}}^{\dagger}=\left(\begin{array}{cccccc}
c_{h,\mathbf{k}\uparrow}^{\dagger} & c_{h,\mathbf{k}\downarrow}^{\dagger} & c_{e_{1},\mathbf{k}\uparrow}^{\dagger} & c_{e_{1},\mathbf{k}\downarrow}^{\dagger} & c_{e_{2},\mathbf{k}\uparrow}^{\dagger} & c_{e_{2},\mathbf{k}\downarrow}^{\dagger}\end{array}\right)
\end{equation}
which allows us to write the fermionic action in the compact form:

\begin{equation}
S=-\int_{k}\Psi_{k}^{\dagger}\hat{\mathcal{G}}_{k}^{-1}\Psi_{k}^ {}+S_{0}\left[M_{i}^{2}\right]
\end{equation}

In the previous expression, $S_{0}\left[M_{i}^{2}\right]$ corresponds
to the terms $M_{i}^{2}$ that arise from the decoupling of the fermionic
interactions. As we explained above, these terms can be ignored for
our purposes. The total Green's function is given by:

\begin{equation}
\hat{\mathcal{G}}_{k}^{-1}=\left(\hat{\mathcal{G}}_{k}^{(0)}\right)^{-1}-\hat{V}_{\mathrm{mag}}-\hat{V}_{iS}-\hat{V}_{C}\label{Greens function}
\end{equation}

The bare part is:

\begin{equation}
\hat{\mathcal{G}}_{k}^{(0)}=\left(\begin{array}{cccccc}
G_{h,k} & 0 & 0 & 0 & 0 & 0\\
0 & G_{h,k} & 0 & 0 & 0 & 0\\
0 & 0 & G_{e_{1},k} & 0 & 0 & 0\\
0 & 0 & 0 & G_{e_{1},k} & 0 & 0\\
0 & 0 & 0 & 0 & G_{e_{2},k} & 0\\
0 & 0 & 0 & 0 & 0 & G_{e_{2},k}
\end{array}\right)\label{G0}
\end{equation}
where $G_{i,k}^{-1}=i\omega_{n}-\xi_{i,\mathbf{k}}$ are the non-interacting
single-particle Green's functions. The interacting parts are:

\begin{equation}
\hat{V}_{\mathrm{mag}}=\left(\begin{array}{cccccc}
0 & 0 & -M_{1,z} & -M_{1,x}+iM_{1,y} & -M_{2,z} & -M_{2,x}+iM_{2,y}\\
0 & 0 & -M_{1,x}-iM_{1,y} & M_{1,z} & -M_{2,x}-iM_{2,y} & M_{2,z}\\
-M_{1,z} & -M_{1,x}+iM_{1,y} & 0 & 0 & 0 & 0\\
-M_{1,x}-iM_{1,y} & M_{1,z} & 0 & 0 & 0 & 0\\
-M_{2,z} & -M_{2,x}+iM_{2,y} & 0 & 0 & 0 & 0\\
-M_{2,x}-iM_{2,y} & M_{2,z} & 0 & 0 & 0 & 0
\end{array}\right)\label{V_SDW}
\end{equation}
and:

\begin{equation}
\hat{V}_{iS}=\left(\begin{array}{cccccc}
0 & 0 & 0 & 0 & 0 & 0\\
0 & 0 & 0 & 0 & 0 & 0\\
0 & 0 & 0 & 0 & i\Delta''_{S,z} & i\left(\Delta''_{S,x}-i\Delta''_{S,y}\right)\\
0 & 0 & 0 & 0 & i\left(\Delta''_{S,x}+i\Delta''_{S,y}\right) & -i\Delta''_{S,z}\\
0 & 0 & -i\Delta''_{S,z} & -i\left(\Delta''_{S,x}-i\Delta''_{S,y}\right) & 0 & 0\\
0 & 0 & -i\left(\Delta''_{S,x}+i\Delta''_{S,y}\right) & i\Delta''_{S,z} & 0 & 0
\end{array}\right)\label{V_SO}
\end{equation}
as well as:

\begin{equation}
\hat{V}_{C}=\left(\begin{array}{cccccc}
0 & 0 & 0 & 0 & 0 & 0\\
0 & 0 & 0 & 0 & 0 & 0\\
0 & 0 & 0 & 0 & -\Delta'_{C} & 0\\
0 & 0 & 0 & 0 & 0 & -\Delta'_{C}\\
0 & 0 & -\Delta'_{C} & 0 & 0 & 0\\
0 & 0 & 0 & -\Delta'_{C} & 0 & 0
\end{array}\right)\label{V_CDW}
\end{equation}

It is now straightforward to integrate out the fermions and obtain
the effective magnetic action:

\begin{equation}
S_{\mathrm{eff}}\left[\mathbf{M}_{1},\mathbf{M}_{2},\boldsymbol{\Delta}''_{S},\Delta'_{C}\right]=-\mathrm{Tr}\ln\left[1-\hat{\mathcal{G}}_{0}\left(\hat{V}_{\mathrm{mag}}+\hat{V}_{iS}+\hat{V}_{C}\right)\right]\approx\sum_{n}\frac{1}{n}\mathrm{Tr}\left[\hat{\mathcal{G}}_{0}\left(\hat{V}_{\mathrm{mag}}+\hat{V}_{iS}+\hat{V}_{C}\right)\right]^{n}
\end{equation}
where, in the last step, we expanded for small $M_{1}$, $M_{2}$.
Here, $\mathrm{Tr}\left(\cdots\right)$ refers to sum over momentum,
frequency and Nambu indices. A straightforward evaluation gives, to
leading order in the coupling between $\boldsymbol{\Delta}''_{S}$,
$\Delta'_{C}$, and $\mathbf{M}_{i}$:

\begin{equation}
S_{\mathrm{eff}}\left[\mathbf{M}_{1},\mathbf{M}_{2},\boldsymbol{\Delta}''_{S},\Delta'_{C}\right]=S\left[\mathbf{M}_{1},\mathbf{M}_{2}\right]+\lambda\boldsymbol{\Delta}''_{S}\cdot\left(\mathbf{M}_{1}\times\mathbf{M}_{2}\right)-\lambda\Delta'_{C}\left(\mathbf{M}_{1}\cdot\mathbf{M}_{2}\right)\label{aux_effective_S-1}
\end{equation}
with the coefficient:

\begin{equation}
\lambda=4\int_{k}G_{h,k}G_{e_{1},k}G_{e_{2},k}\label{lambda_s}
\end{equation}

For perfect nesting, $\delta_{0}=\delta_{2}=0$, this coefficient
vanishes. For a system in proximity to a finite temperature phase
transition, expansion in powers of $\delta_{0}$ gives:

\begin{eqnarray}
\lambda & \approx & 4\rho_{F}T\sum_{n}\int d\xi\frac{1}{\left(i\omega_{n}+\xi\right)}\frac{1}{\left(i\omega_{n}-\xi+\delta_{0}\right)^{2}}\nonumber \\
\lambda & \approx & -8\delta_{0}\rho_{F}T\sum_{n}\int d\xi\frac{1}{\left(i\omega_{n}+\xi\right)}\frac{1}{\left(i\omega_{n}-\xi\right)^{3}}\nonumber \\
\lambda & \approx & -\left(\frac{\delta_{0}}{T}\right)\frac{7\zeta\left(3\right)\rho_{F}}{2\pi^{2}T}\label{ev_lambda_s}
\end{eqnarray}
where $\rho_{F}$ is the density of states at the Fermi level. Therefore,
it is clear that a spin-current density-wave $\boldsymbol{\Delta}''_{S}$
parallel to $\boldsymbol{\varphi}_{\mathrm{SVDW}}$ is triggered by
the SVDW order parameter, $\boldsymbol{\varphi}_{\mathrm{SVDW}}\propto\mathbf{M}_{1}\times\mathbf{M}_{2}$,
whereas a checkerboard charge order $\Delta'_{C}$ is triggered by
the CDW order parameter $\varphi_{\mathrm{CDW}}\propto\mathbf{M}_{1}\cdot\mathbf{M}_{2}$.

\subsection{Non-zero magnetic field}

In the presence of a magnetic field, additional types of fermionic
order are triggered by the condensation of the SVDW and CDW order
parameters. To show that, we first introduce the Zeeman coupling between
the uniform field $\mathbf{H}$ and the electrons:

\begin{equation}
H_{\mathrm{Zeeman}}=\sum_{\mathbf{k},i}\mathbf{H}\cdot\boldsymbol{\sigma}_{\alpha\beta}c_{i,\mathbf{k}\alpha}^{\dagger}c_{i,\mathbf{k}\beta}\label{Zeeman-1}
\end{equation}

We also introduce the $\mathbf{Q}_{1}+\mathbf{Q}_{2}=\left(\pi,\pi\right)$
charge-current density-wave $\Delta''_{C}$ (i.e. a purely imaginary
CDW) and the spin density-wave $\boldsymbol{\Delta}'_{S}$ (i.e. a
purely real SDW) defined by:

\begin{align}
\mathcal{H}_{S} & =\sum_{\mathbf{k}}\boldsymbol{\Delta}'_{S}\cdot\boldsymbol{\sigma}_{\alpha\beta}^{\phantom{\dagger}}\left(c_{e_{2},\mathbf{k}\alpha}^{\dagger}c_{e_{1},\mathbf{k}\beta}^{\phantom{\dagger}}+c_{e_{1},\mathbf{k}\alpha}^{\dagger}c_{e_{2},\mathbf{k}\beta}^{\phantom{\dagger}}\right)\nonumber \\
\mathcal{H}_{iC} & =i\sum_{\mathbf{k}}\Delta''_{C}\delta_{\alpha\beta}\left(c_{e_{2},\mathbf{k}\alpha}^{\dagger}c_{e_{1},\mathbf{k}\beta}^{\phantom{\dagger}}-c_{e_{1},\mathbf{k}\alpha}^{\dagger}c_{e_{2},\mathbf{k}\beta}^{\phantom{\dagger}}\right)\label{eq_CDW_SDW-1}
\end{align}

Following the same steps as in the previous subsection, we obtain
the expanded action:

\begin{equation}
S_{\mathrm{eff}}\left[\mathbf{M}_{1},\mathbf{M}_{2},\boldsymbol{\Delta}'_{S},\Delta''_{C}\right]\approx\sum_{n}\frac{1}{n}\mathrm{Tr}\left[\hat{\mathcal{G}}_{0}\left(\hat{V}_{\mathrm{mag}}+\hat{V}_{S}+\hat{V}_{iC}+\hat{V}_{\mathrm{Zeeman}}\right)\right]^{n}
\end{equation}
where the Nambu-space matrices are given by:
\begin{equation}
\hat{V}_{\mathrm{Zeeman}}=\left(\begin{array}{cccccc}
-H_{z} & -H_{x}+iH_{y} & 0 & 0 & 0 & 0\\
-H_{x}-iH_{y} & H_{z} & 0 & 0 & 0 & 0\\
0 & 0 & -H_{z} & -H_{x}+iH_{y} & 0 & 0\\
0 & 0 & -H_{x}-iH_{y} & h_{z} & 0 & 0\\
0 & 0 & 0 & 0 & -H_{z} & -H_{x}+iH_{y}\\
0 & 0 & 0 & 0 & -H_{x}-iH_{y} & H_{z}
\end{array}\right)\label{V_Zeeman-1}
\end{equation}
and:

\begin{equation}
\hat{V}_{iC}=\left(\begin{array}{cccccc}
0 & 0 & 0 & 0 & 0 & 0\\
0 & 0 & 0 & 0 & 0 & 0\\
0 & 0 & 0 & 0 & i\Delta''_{C} & 0\\
0 & 0 & 0 & 0 & 0 & i\Delta''_{C}\\
0 & 0 & -i\Delta''_{C} & 0 & 0 & 0\\
0 & 0 & 0 & -i\Delta''_{C} & 0 & 0
\end{array}\right)\label{V_c-1}
\end{equation}
as well as:

\begin{equation}
\hat{V}_{S}=\left(\begin{array}{cccccc}
0 & 0 & 0 & 0 & 0 & 0\\
0 & 0 & 0 & 0 & 0 & 0\\
0 & 0 & 0 & 0 & -\Delta'_{S,z} & -\left(\Delta'_{S,x}-i\Delta'_{S,y}\right)\\
0 & 0 & 0 & 0 & -\left(\Delta'_{S,x}+i\Delta'_{S,y}\right) & \Delta'_{S,z}\\
0 & 0 & -\Delta'_{S,z} & -\left(\Delta'_{S,x}-i\Delta'_{S,y}\right) & 0 & 0\\
0 & 0 & -\left(\Delta'_{S,x}+i\Delta'_{S,y}\right) & \Delta'_{S,z} & 0 & 0
\end{array}\right)\label{V_SO-1}
\end{equation}

A straightforward evaluation yields, to leading order in the magnetic
field:

\begin{align}
\mathcal{S}_{\mathrm{eff}} & =\mathcal{S}_{\mathrm{eff}}\left[\mathbf{H}=0\right]+\zeta\left[\left(\mathbf{H}\cdot\mathbf{M}_{1}\right)^{2}+\left(\mathbf{H}\cdot\mathbf{M}_{2}\right)^{2}\right]\label{S_h}\\
 & +\gamma\left[\Delta''_{C}\mathbf{H}\cdot\left(\mathbf{M}_{1}\times\mathbf{M}_{2}\right)+\left(\mathbf{H}\cdot\boldsymbol{\Delta}'_{S}\right)\left(\mathbf{M}_{1}\cdot\mathbf{M}_{2}\right)\right]\nonumber \\
 & +\eta\left[\left(\mathbf{M}_{1}\cdot\mathbf{H}\right)\left(\mathbf{M}_{2}\cdot\boldsymbol{\Delta}'_{S}\right)+\left(\mathbf{M}_{2}\cdot\mathbf{H}\right)\left(\mathbf{M}_{1}\cdot\boldsymbol{\Delta}'_{S}\right)\right]\nonumber 
\end{align}
where we neglected all isotropic biquadratic terms of the form $H^{2}M_{i}^{2}$.
The coefficients are given by:

\begin{eqnarray}
\zeta & = & 4\int_{k}G_{h,k}^{2}G_{e_{1},k}^{2}\nonumber \\
\gamma & = & 4\int_{k}G_{h,k}G_{e_{1},k}G_{e_{2},k}\left(G_{e_{1},k}+G_{e_{2},k}-G_{h,k}\right)\nonumber \\
\eta & = & 4\int_{k}G_{h,k}^{2}G_{e_{1},k}G_{e_{2},k}
\end{eqnarray}

It is useful to perform an expansion around perfect nesting, $\delta_{0}=\delta_{2}=0$.
The coefficients $\zeta$ and $\eta$ become identical in this limit:

\begin{equation}
\zeta=\eta=\frac{\rho_{F}}{T^{2}}\left(\frac{7\zeta\left(3\right)}{2\pi^{2}}\right)
\end{equation}

The fact that $\zeta>0$ implies that the magnetic field induces an
easy plane, rather than an easy axis anisotropy. As for the coefficient
$\eta$, it remains zero in all orders in perturbation theory if an
infinite bandwidth is assumed. However, keeping the top of the hole
pocket $W$ (or bottom of the electron pocket) throughout the calculation
gives:

\begin{equation}
\gamma\approx\frac{\rho_{F}}{T^{2}}\left(\frac{W}{T}\right)^{-2}
\end{equation}

The fact that $\gamma\neq0$ implies that, in the presence of a uniform
field, the SVDW order parameter $\boldsymbol{\varphi}_{\mathrm{SVDW}}\propto\mathbf{M}_{1}\times\mathbf{M}_{2}$
also triggers a charge-current density-wave $\Delta''_{C}$, whereas
the CDW order parameter $\varphi_{\mathrm{CDW}}\propto\mathbf{M}_{1}\cdot\mathbf{M}_{2}$
triggers a spin density-wave of same period, $\boldsymbol{\Delta}'_{S}$.
Although this was expected by symmetry, here we have microscopic expressions
for the corresponding Ginzburg-Landau coefficients. It is interesting
then to compare the coefficient $\gamma$ in Eq. (\ref{S_h}), which
determines the amplitudes of $\Delta''_{C}$ and $\boldsymbol{\Delta}'_{S}$,
to the coefficient $\lambda$ in Eq. (\ref{aux_effective_S-1}), which
determines the amplitudes of $\boldsymbol{\Delta}''_{S}$ and $\Delta'_{C}$.
We find that:

\begin{equation}
\frac{\gamma H}{\lambda}\approx-2.3\left(\frac{T^{2}H}{W^{2}\delta_{0}}\right)
\end{equation}

Therefore, for pnictides whose band dispersions do not deviate strongly
from perfect nesting, and whose bandwidths are not too large either,
it is conceivable that the two coupling constants $\gamma H$ and
$\lambda$ will be of similar order for moderate values of the magnetic
field $H$. As a result, the charge-current density-wave and the spin
density-wave generated in the presence of the field could be as large
as the spin-current density-wave and the charge density-wave generated
in the absence of the field.

\section{Superconducting pairing interactions \label{Appendix_pairing}}

Here we show explicitly that fluctuations associated with an imaginary
SDW instability or with a real CDW instability give rise to attractive
pairing interactions. For our purposes, it is sufficient to consider
only the two Fermi pockets connected by the momentum transfer $\mathbf{Q}=\left(\pi,\pi\right)$
associated with these two ordered states. To simplify the notation,
here we will denote the fermionic operators associated with these
bands by $d_{\mathbf{k}\sigma}$ and $f_{\mathbf{k}\sigma}$. In both
cases, $\mathbf{k}$ is measured relative to the center of each Fermi
pocket. Consider first the action describing the coupling between
the electrons and the complex SDW bosonic field $\boldsymbol{\Delta}_{S}=\Delta_{S}\hat{\mathbf{z}}$
(for simplicity, we consider it polarized along the $z$ axis):

\begin{eqnarray}
S & = & -\int_{k}\left[\left(i\omega_{n}-\varepsilon_{d,\mathbf{k}}\right)d_{\mathbf{k}\sigma}^{\dagger}d_{\mathbf{k}\sigma}+\left(i\omega_{n}-\varepsilon_{f,\mathbf{k}}\right)f_{\mathbf{k}\sigma}^{\dagger}f_{\mathbf{k}\sigma}\right]+g\int_{k,q}\sigma\left(\Delta'_{S,-k-q}d_{\mathbf{k}\sigma}^{\dagger}f_{\mathbf{q}\sigma}+\Delta'_{S,-k-q}f_{\mathbf{k}\sigma}^{\dagger}d_{\mathbf{q}\sigma}\right)\nonumber \\
 &  & +g\int_{k,q}\sigma\left(i\Delta''_{S,-k-q}d_{\mathbf{k}\sigma}^{\dagger}f_{\mathbf{q}\sigma}-i\Delta''_{S,-k-q}f_{\mathbf{k}\sigma}^{\dagger}d_{\mathbf{q}\sigma}\right)\nonumber \\
 &  & +\int_{k}\chi_{S}^{-1}\left(\mathbf{k},\Omega_{n}\right)\Delta'_{S,k}\Delta'_{S,-k}+\int_{k}\chi_{iS}^{-1}\left(\mathbf{k},\Omega_{n}\right)\Delta''_{S,k}\Delta''_{S,-k}\label{action-1}
\end{eqnarray}
where $k\equiv\left(\mathbf{k},\omega_{n}\right)$, $\int_{k}\equiv T\sum\limits _{n}\int\frac{d^{d}k}{\left(2\pi\right)^{d}}$
(with the appropriate bosonic $\Omega_{n}$ or fermionic $\omega_{n}$
Matsubara frequency), and we left implicit the sum over spin indices,
as well as the dependence of the fermionic operators on the fermionic
Matsubara frequencies. $\chi_{S}$ and $\chi_{iS}$ are the susceptibilities
associated with the real and imaginary SDW, and $g$ is the coupling
constant. Note that, because $\mathbf{Q}$ is a commensurate vector,
the real and imaginary SDW fields are independent. Introducing the
four-dimensional Nambu operator:

\begin{equation}
\Psi_{\mathbf{k}}^{\dagger}=\left(\begin{array}{cccc}
d_{\mathbf{k}\uparrow}^{\dagger} & d_{-\mathbf{k}\downarrow} & f_{\mathbf{k}\uparrow}^{\dagger} & f_{-\mathbf{k}\downarrow}\end{array}\right)
\end{equation}
the action can be written conveniently as:

\begin{eqnarray}
S & = & -\int_{k}\Psi_{\mathbf{k}}^{\dagger}\left(i\omega_{n}\hat{1}-\hat{\varepsilon}_{\mathbf{k}}\right)\Psi_{\mathbf{k}}+\int_{k}\chi_{S}^{-1}\left(\mathbf{k},\Omega_{n}\right)\Delta'_{S,k}\Delta'_{S,-k}+\int_{k}\chi_{iS}^{-1}\left(\mathbf{k},\Omega_{n}\right)\Delta''_{S,k}\Delta''_{S,-k}\nonumber \\
 &  & +g\int_{k,q}\Delta'_{S,-k-q}\Psi_{\mathbf{k}}^{\dagger}\hat{\rho}_{S}\Psi_{\mathbf{p}}+g\int_{k,q}\Delta''_{S,-k-q}\Psi_{\mathbf{k}}^{\dagger}\hat{\rho}_{iS}\Psi_{\mathbf{p}}\label{action_Nambu}
\end{eqnarray}
where we defined the $4\times4$ matrices:

\begin{equation}
\hat{\varepsilon}_{\mathbf{k}}=\left(\begin{array}{cc}
\varepsilon_{d,\mathbf{k}}\tau_{z} & 0\\
0 & \varepsilon_{f,\mathbf{k}}\tau_{z}
\end{array}\right)\:;\;\;\hat{\rho}_{S}=\left(\begin{array}{cc}
0 & \tau_{0}\\
\tau_{0} & 0
\end{array}\right)\equiv\tau_{0}\otimes\sigma_{x}\:;\;\;\hat{\rho}_{iS}=\left(\begin{array}{cc}
0 & i\tau_{z}\\
-i\tau_{z} & 0
\end{array}\right)\equiv-\tau_{z}\otimes\sigma_{y}\label{matrix_elements}
\end{equation}
where $\tau_{i}$ are the Pauli matrices and $0$ denotes the $2\times2$
matrix whose elements are all zero. To obtain the Eliashberg-like
gap equations, we need to solve Dyson's equation:

\begin{equation}
\hat{G}_{k}^{-1}=\hat{G}_{0,k}^{-1}-\hat{\Sigma}_{k}\label{Dyson}
\end{equation}
with $\hat{G}_{0,k}^{-1}=i\omega_{n}\hat{1}-\hat{\varepsilon}_{\mathbf{k}}$
and the one-loop self-energy:

\begin{equation}
\hat{\Sigma}_{k}=g^{2}\int_{q}\chi_{S}\left(k-q\right)\hat{\rho}_{S}\hat{G}_{q}\hat{\rho}_{S}+g^{2}\int_{q}\chi_{iS}\left(k-q\right)\hat{\rho}_{iS}\hat{G}_{q}\hat{\rho}_{iS}\label{self_energy}
\end{equation}

It is convenient to parametrize the self-energy by:

\begin{equation}
\hat{\Sigma}_{k}=\left(\hat{1}-\hat{Z}_{k}\right)i\omega_{n}+\hat{W}_{k}+\hat{\xi}_{k}\label{self_energy_ansatz}
\end{equation}
where we introduced the imaginary normal components $Z_{\mu,k}$,
the real normal components $\xi_{\mu,k}$, and the anomalous components
$W_{\mu,k}$ ($\mu=d,f$ is a band index):

\begin{equation}
\hat{Z}_{k}=\left(\begin{array}{cc}
Z_{d,k}\tau_{0} & 0\\
0 & Z_{f,k}\tau_{0}
\end{array}\right)\:;\;\;\hat{W}_{k}=\left(\begin{array}{cc}
W_{d,k}\tau_{x} & 0\\
0 & W_{f,k}\tau_{x}
\end{array}\right)\:;\;\;\hat{\xi}_{k}=\left(\begin{array}{cc}
\xi_{d,k}\tau_{z} & 0\\
0 & \xi_{f,k}\tau_{z}
\end{array}\right)\label{aux_self_energy}
\end{equation}

The superconducting gap in band $\mu$ is therefore proportional to
$W_{\mu,k}$. Using Eqs. (\ref{Dyson}) and (\ref{self_energy_ansatz}),
it is straightforward to invert the matrix and obtain $\hat{G}$.
Substituting it in (\ref{self_energy}) and comparing back with Eq.
(\ref{self_energy_ansatz}), we arrive at a set of six self-consistent
equations. Four of them have the same form for either real or imaginary
SDW, namely, the two equations that renormalize the dispersion $\tilde{\varepsilon}_{a,k}=\xi_{a,k}+\varepsilon_{\mathbf{k}}$
and the two that renormalize the quasi-particle weights $Z_{\mu,\mathbf{k}}$.
However, the two self-consistent gap equations acquire different forms:

\begin{eqnarray}
W_{d,k} & = & -\int_{q}\left[g^{2}\chi_{S}\left(k-q\right)\right]\frac{W_{f,q}}{D_{f,q}}-\int_{q}\left[-g^{2}\chi_{iS}\left(k-q\right)\right]\frac{W_{f,q}}{D_{f,q}}\nonumber \\
W_{f,k} & = & -\int_{q}\left[g^{2}\chi_{S}\left(k-q\right)\right]\frac{W_{d,q}}{D_{d,q}}-\int_{q}\left[-g^{2}\chi_{iS}\left(k-q\right)\right]\frac{W_{d,q}}{D_{d,q}}
\end{eqnarray}
where we defined $D_{\mu,q}^{2}=Z_{\mu,q}^{2}\omega_{n}^{2}+\tilde{\varepsilon}_{\mu,q}^{2}+W_{\mu,q}^{2}$.
From the form of these equations, it becomes clear that while the
fluctuations near the real SDW instability give rise to a repulsive
inter-band pairing interaction, $V_{df}\propto g^{2}\chi_{S}$, the
fluctuations near the imaginary SDW instability promote an attractive
inter-band pairing interaction, $V_{df}\propto-g^{2}\chi_{iS}$. This
difference relies ultimately on the different structures of the matrix
elements (\ref{matrix_elements}) in Nambu space.

A similar analysis can be performed in the charge channel:

\begin{eqnarray}
S & = & -\int_{k}\left[\left(i\omega_{n}-\varepsilon_{d,\mathbf{k}}\right)d_{\mathbf{k}\sigma}^{\dagger}d_{\mathbf{k}\sigma}+\left(i\omega_{n}-\varepsilon_{f,\mathbf{k}}\right)f_{\mathbf{k}\sigma}^{\dagger}f_{\mathbf{k}\sigma}\right]+g\int_{k,q}\left(\Delta'_{C,-k-q}d_{\mathbf{k}\sigma}^{\dagger}f_{\mathbf{q}\sigma}+\Delta'_{C,-k-q}f_{\mathbf{k}\sigma}^{\dagger}d_{\mathbf{q}\sigma}\right)\nonumber \\
 &  & +g\int_{k,q}\left(i\Delta''_{C,-k-q}d_{\mathbf{k}\sigma}^{\dagger}f_{\mathbf{q}\sigma}-i\Delta''_{C,-k-q}f_{\mathbf{k}\sigma}^{\dagger}d_{\mathbf{q}\sigma}\right)\nonumber \\
 &  & +\int_{k}\chi_{C}^{-1}\left(\mathbf{k},\Omega_{n}\right)\Delta'_{C,k}\Delta'_{C,-k}+\int_{k}\chi_{iC}^{-1}\left(\mathbf{k},\Omega_{n}\right)\Delta''_{C,k}\Delta''_{C,-k}\label{action_C}
\end{eqnarray}

In Nambu space, we obtain:

\begin{eqnarray}
S & = & -\int_{k}\Psi_{\mathbf{k}}^{\dagger}\left(i\omega_{n}\hat{1}-\hat{\varepsilon}_{\mathbf{k}}\right)\Psi_{\mathbf{k}}+\int_{k}\chi_{C}^{-1}\left(\mathbf{k},\Omega_{n}\right)\Delta'_{C,k}\Delta'_{C,-k}+\int_{k}\chi_{iC}^{-1}\left(\mathbf{k},\Omega_{n}\right)\Delta''_{C,k}\Delta''_{C,-k}\nonumber \\
 &  & +g\int_{k,q}\Delta'_{C,-k-q}\Psi_{\mathbf{k}}^{\dagger}\hat{\rho}_{C}\Psi_{\mathbf{p}}+g\int_{k,q}\Delta''_{C,-k-q}\Psi_{\mathbf{k}}^{\dagger}\hat{\rho}_{iC}\Psi_{\mathbf{p}}\label{action_Nambu_C}
\end{eqnarray}

where we defined the $4\times4$ matrices:

\begin{equation}
\hat{\rho}_{C}=\left(\begin{array}{cc}
0 & \tau_{z}\\
\tau_{z} & 0
\end{array}\right)\equiv\tau_{z}\otimes\sigma_{x}\:;\;\;\hat{\rho}_{iC}=\left(\begin{array}{cc}
0 & i\tau_{0}\\
-i\tau_{0} & 0
\end{array}\right)\equiv-\tau_{0}\otimes\sigma_{y}\label{matrix_elements_C}
\end{equation}

Solving the one-loop Dyson equation, we obtain the two self-consistent
gap equations:

\begin{eqnarray}
W_{d,k} & = & -\int_{q}\left[-g^{2}\chi_{C}\left(k-q\right)\right]\frac{W_{f,q}}{D_{f,q}}-\int_{q}\left[g^{2}\chi_{iC}\left(k-q\right)\right]\frac{W_{f,q}}{D_{f,q}}\nonumber \\
W_{f,k} & = & -\int_{q}\left[-g^{2}\chi_{C}\left(k-q\right)\right]\frac{W_{d,q}}{D_{d,q}}-\int_{q}\left[g^{2}\chi_{iC}\left(k-q\right)\right]\frac{W_{d,q}}{D_{d,q}}
\end{eqnarray}

Therefore, in the charge channel, real CDW fluctuations promote inter-band
pairing attraction, $V_{df}\propto-g^{2}\chi_{C}$, whereas imaginary
CDW fluctuations promote repulsion, $V_{df}\propto g^{2}\chi_{iC}$.

\end{widetext}

\begin{thebibliography}{10}
\bibitem{reviews} K. Ishida, Y. Nakai and H. Hosono, J. Phys. Soc.
Japan \textbf{78}, 062001 (2009); D. C. Johnston, Adv. Phys. \textbf{59},
803 (2010); J. Paglione and R. L. Greene, Nature Phys. \textbf{6},
645 (2010); P. C. Canfield and S. L. Bud'ko, Annu. Rev. Cond. Mat.
Phys. \textbf{1}, 27 (2010); H. H. Wen and S. Li, Annu. Rev. Cond.
Mat. Phys. \textbf{2}, 121 (2011).

\bibitem{Dagotto12} P. Dai, J. Hu, and E. Dagotto, Nature Phys. \textbf{8},
709 (2012).

\bibitem{Fang08} C. Fang, H. Yao, W.-F. Tsai, J. P. Hu and S. A.
Kivelson, Phys. Rev. B \textbf{77}, 224509 (2008)

\bibitem{Xu08} C. Xu, M. Muller, and S. Sachdev, Phys. Rev. B \textbf{78},
020501(R) (2008).

\bibitem{Johannes09} M. D. Johannes and I. I. Mazin, Phys. Rev. B
\textbf{79}, 220510(R) (2009).

\bibitem{Eremin10} I. Eremin and A. V. Chubukov, Phys. Rev. B \textbf{81},
024511 (2010).

\bibitem{Abrahams11} E. Abrahams and Q. Si, J. Phys.: Condens. Matter
23, 223201 (2011).

\bibitem{Fernandes12} R. M. Fernandes, A. V. Chubukov, J. Knolle,
I. Eremin and J. Schmalian, Phys. Rev. B \textbf{85}, 024534 (2012).

\bibitem{Dagotto14} S. Liang, A. Mukherjee, N. D. Patel, E. Dagotto,
and A. Moreo, Phys. Rev. B \textbf{90}, 184507 (2014).

\bibitem{Kivelson14} L. Nie, G. Tarjus, and S. A. Kivelson, PNAS
\textbf{111}, 7980 (2014).

\bibitem{Fernandes14} R. M. Fernandes, A. V. Chubukov, and J. Schmalian,
Nature Phys. \textbf{10}, 97 (2014).

\bibitem{Fisher10} J.-H. Chu, J. G. Analytis, K. De Greve, P. L.
McMahon, Z. Islam, Y. Yamamoto, and I. R. Fisher, Science \textbf{329},
824 (2010).

\bibitem{Davis10} T.-M. Chuang, M. P. Allan, J. Lee, Y. Xie, N. Ni,
S. L. Bud'ko, G. S. Boebinger, P. C. Canfield, and J. C. Davis, Science
\textbf{327}, 181 (2010).

\bibitem{ZXshen11} M. Yi, D. Lu, J.-H. Chu, J. G. Analytis, A. P.
Sorini, A. F. Kemper, B. Moritz, S.-K. Mo, R. G. Moore, M. Hashimoto,
W. S. Lee, Z. Hussain, T. P. Devereaux, I. R. Fisher, Z.-X. Shen,
Proc. Nat. Acad. Sci. 2011 \textbf{108}, 6878 (2011).

\bibitem{Fisher12} J.-H. Chu, H.-H. Kuo, J. G. Analytis, and I. R.
Fisher, Science \textbf{337}, 710 (2012).

\bibitem{Matsuda12} S. Kasahara, H. J. Shi, K. Hashimoto, S. Tonegawa,
Y. Mizukami, T. Shibauchi, K. Sugimoto, T. Fukuda, T. Terashima, A.
H. Nevidomskyy, and Y. Matsuda, Nature \textbf{486}, 382 (2012).

\bibitem{Gallais13} Y. Gallais, R. M. Fernandes, I. Paul, L. Chauviere,
Y.-X. Yang, M.-A. Measson, M. Cazayous, A. Sacuto, D. Colson, and
A. Forget, Phys. Rev. Lett. \textbf{111}, 267001 (2013).

\bibitem{Dai14} X. Lu, J. T. Park, R. Zhang, H. Luo, A. H. Nevidomskyy,
Q. Si, and P. Dai, Science \textbf{345}, 657 (2014).

\bibitem{Rosenthal14} E. P. Rosenthal, E. F. Andrade, C. J. Arguello,
R. M. Fernandes, L. Y. Xing, X. C. Wang, C. Q. Jin, A. J. Millis,
and A. N. Pasupathy, Nature Phys. \textbf{10}, 225 (2014).

\bibitem{w_ku10} C. C. Lee, W. G. Yin, and W. Ku, Phys. Rev. Lett.
\textbf{103}, 267001 (2009).

\bibitem{Devereaux12} R. Applegate, R. R. P. Singh, C.-C. Chen, and
T. P. Devereaux, Phys. Rev. B \textbf{85}, 054411 (2012).

\bibitem{Fernandes_Millis} R. M. Fernandes and A. J. Millis, Phys.
Rev. Lett. \textbf{111}, 127001 (2013).

\bibitem{Lederer14} S. Lederer, Y. Schattner, E. Berg, and S. A.
Kivelson, Phys. Rev. Lett. \textbf{114}, 097001 (2015).

\bibitem{Kang14} J. Kang, A. F. Kemper, and R. M. Fernandes, Phys.
Rev. Lett. \textbf{113}, 217001 (2014).

\bibitem{Kim10} M. G. Kim, A. Kreyssig, A. Thaler, D. K. Pratt, W.
Tian, J. L. Zarestky, M. A. Green, S. L. Bud'ko, P. C. Canfield, R.
J. McQueeney, and A. I. Goldman, Phys. Rev. B \textbf{82}, 220503(R)
(2010)

\bibitem{Avci14} S. Avci, O. Chmaissem, J. M. Allred, S. Rosenkranz,
I. Eremin, A. V. Chubukov, D. E. Bulgaris, D. Y. Chung, M. G. Kanatzidis,
J.-P Castellan, J. A. Schlueter, H. Claus, D. D. Khalyavin, P. Manuel,
A. Daoud-Aladine, and R. Osborn, Nature Comm. \textbf{5}, 3845 (2014)

\bibitem{Bohmer14} A. E. Böhmer, F. Hardy, L. Wang, T. Wolf, P. Schweiss,
and C. Meingast, arXiv:1412.7038.

\bibitem{xiaoyu14} X. Wang and R. M. Fernandes, Phys. Rev. B \textbf{89},
144502 (2014)

\bibitem{Wang_arxiv_14} X. Wang, J. Kang, and R. M. Fernandes, Phys.
Rev. B \textbf{91}, 024401 (2015).

\bibitem{Kang15} J. Kang, X. Wang, A. V. Chubukov, and R. M. Fernandes,
Phys. Rev. B \textbf{91}, 121104(R) (2015).

\bibitem{Gastiasoro15} M. N. Gastiasoro and B. M. Andersen, arXiv:1502.05859.

\bibitem{Lorenzana08} J. Lorenzana, G. Seibold, C. Ortix, and M.
Grilli, Phys. Rev. Lett. \textbf{101}, 186402 (2008).

\bibitem{giovannetti} G. Giovannetti, C. Ortix, M. Marsman, M. Capone,
J. Brink and J. Lorenzana, Nat. Comm. 2, \textbf{398} (2011).

\bibitem{Brydon11} P. M. R. Brydon, J. Schmiedt, and C. Timm, Phys.
Rev. B \textbf{84}, 214510 (2011).

\bibitem{Chern12} G.-W. Chern, R. M. Fernandes, R. Nandkishore, and
A. V. Chubukov, Phys. Rev. B \textbf{86}, 115443 (2012).

\bibitem{Chakravarty01} S. Chakravarty, R. B. Laughlin, D. K. Morr,
and C. Nayak, Phys. Rev. B \textbf{63}, 094503 (2001).

\bibitem{chandra} P. Chandra, P. Coleman and A. I. Larkin, Phys.
Rev. Lett. \textbf{64}, 88-91 (1990).

\bibitem{Berg10} E. Berg, S. A. Kivelson, and D. J. Scalapino, Phys.
Rev. B \textbf{81}, 172504 (2010).

\bibitem{Batista11} Y. Kamiya, N. Kawashima, and C. D. Batista, Phys.
Rev. B \textbf{84}, 214429 (2011).

\bibitem{footnote-symmetries} Note that, in contrast to the stripe
and CSDW phases, which are collinear and display a residual $O(2)$
symmetry, the non-collinear SVC phase breaks completely the spin-rotational
symmetry.

\bibitem{Batista09} K. A. Al-Hassanieh, C. D. Batista, G. Ortiz,
and L. N. Bulaevskii, Phys. Rev. Lett. \textbf{103}, 216402 (2009).

\bibitem{footnote-inversion} Strictly speaking, the SVDW phase retains
a mirror symmetry, and is therefore not chiral. Instead, the SVDW
phase breaks inversion symmetry. We will nevertheless use the term
``chiral'' to describe it, as it captures the ``handness'' of
the spins around a plaquette.

\bibitem{Matan09} K. Matan, R. Morinaga, K. Iida, and T. J. Sato,
Phys. Rev. B \textbf{79}, 054526 (2009).

\bibitem{Dai13} H. Luo, M. Wang, C. Zhang, L.-P. Regnault, R. Zhang,
S. Li, J. Hu, and P. Dai, Phys. Rev. Lett. 111, 107006 (2013).

\bibitem{Tucker14} G. S. Tucker, R. M. Fernandes, D. K. Pratt, A.
Thaler, N. Ni, K. Marty, A. D. Christianson, M. D. Lumsden, B. C.
Sales, A. S. Sefat, S. L. Bud'ko, P. C. Canfield, A. Kreyssig, A.
I. Goldman, and R. J. McQueeney, Phys. Rev. B \textbf{89}, 180503(R)
(2014).

\bibitem{Christensen15}  M. H. Christensen, Jian Kang, B. M. Andersen,
I. Eremin, and R. M. Fernandes, arXiv:1508.01763 (2015).

\bibitem{Korshunov02} S. E. Korshunov, Phys. Rev. Lett. \textbf{88},
167007 (2002).

\bibitem{Vicari05} M. Hasenbusch, A. Pelissetto, and E. Vicari, J.
Stat. Mech. P12002 (2005).

\bibitem{reviews_pairing} P. J. Hirschfeld, M. M. Korshunov, and
I. I. Mazin, Rep. Prog. Phys. \textbf{74}, 124508 (2011); A. V. Chubukov,
Annu. Rev. Cond. Mat. Phys. \textbf{3}, 57 (2012).

\bibitem{Fujimoto11} S. Fujimoto Phys. Rev. Lett. \textbf{106}, 196407
(2011).

\bibitem{Chakravarty14} C.-H. Hsu and S. Chakravarty, Phys. Rev.
B \textbf{90}, 134507 (2014).

\bibitem{Blumberg15} H.-H. Kung, R. E. Baumbach, E. D. Bauer, V.
K. Thorsmolle1, W.-L. Zhang, K. Haule, J. A. Mydosh, and G. Blumberg,
Science \textbf{347}, 1339 (2015).

\bibitem{Allred15} J. M. Allred \emph{et al}, arXiv:1505.06175 (2015).

\bibitem{Mallet15} B. P. P. Mallett, Yu. G. Pashkevich, A. Gusev,
T. Wolf, and C. Bernhard, arXiv:1506.00786 (2015).

\bibitem{Nagaosa07} S. Oneda and N. Nagaosa, Phys. Rev. Lett. \textbf{99},
027206 (2007).

\end{thebibliography}
\end{document}